\documentclass{article}
\usepackage[utf8]{inputenc}

\usepackage{color}
\usepackage{amsmath}
\usepackage{tabularx}
\usepackage{amsfonts}
\usepackage{graphicx}
\usepackage[dvipsnames]{xcolor}
\definecolor{refs}{RGB}{245,156,74}
 \usepackage[colorlinks=cyan,hyperfootnotes=true,citecolor=cyan]{hyperref}
\usepackage{footnote}
\usepackage{footmisc}
\makesavenoteenv{tabular}
\makesavenoteenv{table}

\renewcommand{\thefootnote}{\alph{footnote}}

\newcommand{\astfootnote}[1]{%
\let\oldthefootnote=\thefootnote%
\setcounter{footnote}{0}%
\renewcommand{\thefootnote}{\fnsymbol{footnote}}%
\footnote{#1}%
\let\thefootnote=\oldthefootnote%
}

\renewcommand{\thefootnote}{\alph{footnote}}

\newcommand{\dagfootnote}[1]{%
\let\oldthefootnote=\thefootnote%
\setcounter{footnote}{1}%
\renewcommand{\thefootnote}{\fnsymbol{footnote}}%
\footnote{#1}%
\let\thefootnote=\oldthefootnote%
}

\renewcommand{\thefootnote}{\alph{footnote}}

\newcommand{\thirdfootnote}[1]{%
\let\oldthefootnote=\thefootnote%
\setcounter{footnote}{2}%
\renewcommand{\thefootnote}{\fnsymbol{footnote}}%
\footnote{#1}%
\let\thefootnote=\oldthefootnote%
}

\newcommand{\be}{\begin{equation}}
\newcommand{\ee}{\end{equation}}
\newcommand{\bea}{\begin{eqnarray}}
\newcommand{\eea}{\end{eqnarray}}

\newcommand{\tetrad}{\theta}
\newcommand{\cotetrad}{e}

\newcommand{\spinconnection}{\omega}

\newcommand{\Lorentz}{\Lambda}

\newcommand{\lapse}{\alpha}
\newcommand{\shift}{\beta}
\newcommand{\inducedmetric}{\gamma}
\newcommand{\normalvector}{\xi}
\newcommand{\momenta}{\pi}

\title{Hamilton's equations in the covariant teleparallel\\ equivalent of general relativity}

\author{Laxmipriya Pati${}^{1,3}$\astfootnote{\href{mailto:laxmipriya.pati@ut.ee}{laxmipriya.pati@ut.ee}},  Daniel Blixt$^{2}$\dagfootnote{\href{mailto:danielkristoffer.blixt-ssm@unina.it}{danielkristoffer.blixt-ssm@unina.it}} \ and  Mar\'ia-Jos\'e Guzm\'an${}^{3}$\small \thirdfootnote{\href{mailto:mjguzman@ut.ee}{mjguzman@ut.ee}}, 
\\ 
{\small ${}^1$ {\it Department of Mathematics, Birla Institute of Technology and Science-Pilani, Hyderabad Campus}}\\
{\small {\it Hyderabad-500078, India}}\\
{\small $^2$ {\it Scuola Superiore Meridionale}}\\
{\small {\it Largo S. Marcellino 10, I-80138 Napoli, Italy}}\\
{\small $^3$ {\it Laboratory of Theoretical Physics, Institute of Physics, University of Tartu}}\\
{\small {\it W. Ostwaldi 1, Tartu 50411, Estonia}}
}

\begin{document}

\maketitle

\begin{abstract}
We present Hamilton's equations for the teleparallel equivalent of general relativity (TEGR), which is a reformulation of general relativity based on a curvatureless, metric compatible, and torsionful connection. For this, we consider the Hamiltonian for TEGR obtained through the vector, antisymmetric, symmetric and trace-free, and trace irreducible decomposition of the phase space variables. We present the Hamiltonian for TEGR in the covariant formalism for the first time in the literature, by considering a spin connection depending on Lorentz matrices. We introduce the mathematical formalism necessary to compute Hamilton's equations  in both Weitzenb\"{o}ck gauge and covariant formulation, where for the latter we must introduce new fields: Lorentz matrices and their associated momenta. We also derive explicit relations between the conjugate momenta of the tetrad and the conjugate momenta for the metric that are traditionally defined in GR, which are important to compare both formalisms.
\end{abstract}

\section{Introduction}

The success of Einstein's  theory of general relativity (GR) has constantly been confirmed over the years, with one of its earliest triumphs being able to provide the correct prediction for the bending of light by the Sun  \cite{Dyson19}. Nowadays the most recent groundbreaking observations prove the predictions of GR correct with the observation of gravitational waves from a binary black hole merger  \cite{LIGOScientific:2016aoc}, complimented with the simultaneous detection of both light and gravitational waves in a binary neutron star inspiral \cite{LIGOScientific:2017vwq} which have opened a new window for multimessenger astronomy. GR also allows the inclusion of a cosmological constant $\Lambda$ used to explain the late accelerated expansion of the Universe. Thus, GR encodes the  fundamental cosmological knowledge through the current standard model for cosmology, the so-called $\Lambda$CDM model, for its main components are the cosmological constant and cold dark matter.\\

Despite this success, GR is a theory that still has many elusive open questions \cite{Shankaranarayanan:2022wbx}. To start with, GR is unable to explain the smallness of the cosmological constant, which, corresponding to the value of the vacuum energy density, is predicted to be 120 orders of magnitude larger by quantum field theory \cite{Burgess04}. Therefore, in order to explain the accelerated expansion of the Universe, physicists resort to the concept of dark energy, a component with mysterious properties that has not been observed directly but is predicted by several modifications to GR. Moreover, GR cannot be described as a quantum field theory in the same way as the other fundamental forces are, hence it cannot be directly incorporated into the standard model of particle physics. Among other problems of GR are the tensions in cosmological data such as the discrepancy in the measurement of the Hubble parameter at late and early times \cite{Aghanim20,Riess18}. Finally, the strong evidence for inflation \cite{Bertone05,Perlmutter99,Peebles03,Clowe06,Riess98,Corbelli00,Weinberg89,Copeland06} contrasts with the lack of theoretical tools needed to describe it, since the hypothetical inflaton field has not been discovered. To explain the above issues one can introduce additional fields that would be responsible for these large-scale differences while retaining the well-observed short-distance predictions of general relativity. In this method, it is assumed that GR can be modified or expanded, and such differences could explain cosmological observations. Modifications to GR should be also consistent with standard solar systems tests. It is generally the case, however, that the gauge symmetry of GR is broken in such modifications, which leads to the creation of new degrees of freedom \cite{Harko18,Capozziello11,Xu19}.\\

General relativity is a classical theory for a massless spin-2 field. It is described by Einstein's field equation and it is Lorentz and diffeomorphism invariant. We know that such equations are obtained applying the variational principle in Hilbert's  action formulation, generally known as the Einstein-Hilbert (EH) action. The equations obtained from it fully satisfy these symmetries and lead to Einstein's field equations. This action is formulated in terms of the Ricci scalar, which is built from the Levi-Civita connection that is metric compatible, has curvature and it is torsion free. It is less known that, alternatively to this connection, (while still assuming vanishing nonmetricity) we can use the curvature-free Weitzenb\"{o}ck connection to build the covariant derivative, which defines the teleparallel framework. In this way we can  describe the effects of gravitation in terms of torsion rather than curvature. Analogously, we can choose to work with a connection with purely nonmetricity and vanishing curvature and torsion, and we again obtain an equivalent theory with the same dynamics dictated by Einstein's equations. We can formulate three actions:  the EH action, the teleparallel equivalent of general relativity (TEGR) action, and the symmetric teleparallel equivalent of general relativity (STEGR) action, defined in terms of Lagrangians built from the Ricci scalar $\mathbb R$, the torsion scalar $\mathbb T$ and the nonmetricity scalar $\mathbb Q$. This trio renders Einstein's equations; therefore, all are  classically equivalent and possess its same well-known cosmological and black hole solutions, for example. The Lagrangians differ among each other by boundary terms, which do not affect the dynamics of the equations of motion, therefore the three theories have the same number of degrees of freedom. They are incidentally referred to as the ``geometrical trinity of gravity" \cite{Jimenez19}, and are the foundational blocks for building modifications to gravity, since when taking nonlinear functions of the scalars, we obtain $f(\mathbb T)$ and $f(\mathbb Q)$ theories of modified gravity that have different equations of motion and more degrees of freedom.\\

The EH action classically has to be supplemented  with a boundary term that does not change the field equations. This is made through the incorporation of the York-Gibbons-Hawking (YGH) boundary term \cite{York72,Gibbons77}, which needs to be considered for study of physics in the boundary of a manifold. Such boundary term encapsulates the terms in the EH action that contain second-order derivatives of the metric. Einstein's noncovariant formulation highlights the importance of setting the coordinates, which is translated into the fixation of the gauge in the new version of GR from a modern perspective. 
This term is also important in order to define the gravitational energy-momentum covariantly, a “background structure” must be introduced so that the theory can be “covariantized”, such as auxiliary reference metrics \cite{Rosen40} or auxiliary reference connections \cite{Tomboulis17}. The boundary conditions for the dynamical fields \cite{Deser19} are sufficient to provide asymptotically symmetric solutions for specific cases. 
The condition of the inclusion of the YGH boundary term is in TEGR and STEGR replaced by a gauge condition. In \cite{Gomes:2022vrc} the condition ``inertial frame'' have been shown to give the expected result and it is here argued that this approach is not \textit{ad hoc} in contrast to the introduction of the YGH boundary term.\\
 
The Hamiltonian formalism and Dirac's algorithm for constrained Hamiltonian systems is one of the most important theoretical tools to study any physical theory. It facilitates the identification of physical degrees of freedom by classifying constraints in first and second class. First class constraints are associated with gauge symmetries of the theory, so each one of them is associated with a trivial degree of freedom. Second class constraints can be grouped as pairs of spurious canonical variables. This formalism is essential for canonical quantization of gravity, and its application to GR is an important theoretical landmark. All equations of motion that originate from a variational principle in the Lagrangian formulation can be represented analogously in the Hamiltonian formalism through Hamilton's equations, which are essentially dynamical equations for the position variable $q_{i}$ and its conjugate momenta $p^{i}$. The pioneering research of Arnowitt, Deser, and Misner (ADM) \cite{Arnowitt59}, where the Hamiltonian formulation of GR was derived, currently constitutes the cornerstone of numerical relativity, and has significantly advanced our understanding of gravitation and its highly nonlinear dynamics. By utilizing the ADM formalism it is possible to express GR action in 3+1 decomposition. This approach foliates spacetime into a group of spacelike hypersurfaces $\Sigma_{t}$, by splitting the metric $g_{\mu\nu}$ into lapse $\alpha$ and shift functions $\beta^{i}$, and a three-dimensional spatial metric denoted by  $\gamma_{ij}$. The dynamics of GR are encoded in the spatial metric and its conjugate momenta, while lapse and shift are introduced into the Hamiltonian formulation as Lagrange multipliers. These parameters enable the transformation of Einstein's field equations into the 3+1 decomposition. Since Hamilton's equations are a set of first order differential equations they can also be used in the study of dynamical systems. As of now the standard Hamiltonian analysis in GR \cite{Arnowitt59} is done for an EH action supplemented by the YGH boundary term. In this work instead we will analyze GR through the TEGR action formulation, which differs from the EH action by a total derivative of the trace of the torsion tensor. Note that this boundary term is different from the YGH term. Nevertheless, it gives the expected result when choosing the ``inertial frame'' gauge \cite{Gomes:2022vrc}. Since TEGR differs by a boundary term from the standard Hamiltonian formulation of GR, it is expected that this will be reflected in Hamilton's equations. In particular, the difference is expected to depend on the torsion tensor since the boundary term itself depends on it.\\
 
Hamilton's equations determine the time evolution of the canonical variables, which is crucial to assess the well posedness of the Cauchy problem, as well as to determine the behavior of the degrees of freedom in the theory. In general relativity, they are the Hamiltonian equivalent of the 3+1 decomposition of the Lagrange equations of motion, and they determine the time evolution of the induced metric and its time derivative (or momenta, in the Hamiltonian picture). The aim of this work is to compute Hamilton's equations in the teleparallel equivalent of general relativity, in order to better understand the dynamical behavior of the theory and compare it with the GR case. The Hamiltonian analysis for TEGR has been developed in the Weitzenb\"{o}ck gauge and, up to now, was never performed by using the covariant formulation, although some partial results can be found in \cite{Blixt19a,Blixt19,Golovnev:2021omn}. However, to our knowledge, an explicit derivation of Hamilton's equations has only been presented in an old Master thesis \cite{MTh} and partially done in \cite{Blagojevic00}. In this work we present a closed form for the kinematic Hamiltonian for covariant TEGR, which can be found also in \cite{thesis} for new general relativity and $f(\mathbb T)$ gravity. For this, it is also necessary to extend the phase space by including Lorentz matrices defining a metric teleparallel spin connection. Rather than the 10 elements of a metric, the field variables in teleparallel gravity are composed of 16 components of a tetrad. On top of that, six extra fields have to be added, which are related to local Lorentz transformations, but they are removed by the introduction of six additional primary constraints, therefore they do not represent additional degrees of freedom and are pure gauge. Our aim is that the present study of the 3+1 decomposition of the equations of motion of TEGR opens the stage for the study of numerical relativity in the tetrad formalism. We expect that our work can set the basis for the study of strong hyperbolicity in TEGR, which is essential for implementing stable numerical codes in numerical relativity \cite{Gourgoulhon:2007ue,Baumgarte:2010ndz}. This field of research is essential for the description of physical phenomena in the strong gravity regime, such as the merger of binary pairs of black holes and/or neutron stars. However, the role of the tetrad formalism and moreover, the spin connection, for achieving strong hyperbolicity, is far from being studied, therefore providing an additional motivation for this work. In this work we do not consider nonlinear modifications to TEGR, but it is straightforward to extend our work to modified teleparallel theories. In the future, our work could be used to resolve controversies regarding the degrees of freedom in $f(\mathbb T)$ gravity \cite{Li:2011rn,Ferraro:2018tpu,Ferraro:2018axk,Blagojevic:2020dyq,Golovnev:2020nln,Golovnev:2020zpv,Blixt:2020ekl}.\\

The outline of this paper is as follows. In Sec. \ref{sec:tel} we  introduce the basic mathematical formalism for the teleparallel equivalent of general relativity considering both the tetrad and the spin connection as dynamical variables, and we introduce the foundations for the 3+1 decomposition in the tetrad. Based on this, in Sec. \ref{sec:Ham} we introduce the irreducible decomposition of the conjugate momenta, with which we compute the Hamiltonian of TEGR in the covariant formulation and in the Weitzenb\"{o}ck gauge. Sec. \ref{sec:HeqsW}  is devoted to computing Hamilton's equations of TEGR by taking zero spin connection. In Sec. \ref{sec:Heqscov} we present the computation of Hamilton's equation for the covariant formulation of TEGR. In Sec. \ref{sec:disc} we discuss our findings by comparing them with previous works in the literature. Lastly in Sec. \ref{sec:conc} we summarize our results. Additionally, we provide some useful identities in Appendix \ref{app:A}, we review Hamilton's equations of GR in Appendix \ref{app:B}, and provide a comparison of our results with \cite{Blagojevic00} in Appendix \ref{app:C}. 

\section{Teleparallel gravity and tetrad 3+1 decomposition}
\label{sec:tel}

Throughout this work, we will use the sign convention for the Minkowski metric as the mostly positive one, i.e. $\eta_{AB}=\mathrm{diag}(-1,1,1,1)$. Greek letters $\mu,\nu,\rho,\ldots$ denote spacetime indices and lowercase Latin letters $i=1,2,3$ indicate spatial indices. Lorentz tangent space indices are denoted by the uppercase first letters of the Latin alphabet $A,B,C,\ldots$, and their spatial part is denoted with hats $\hat{A},\hat{B},\hat{C},\ldots$. We consider a field of tetrads on each point of spacetime with components $\tetrad^A{}_{\mu}$ and the components of the inverse tetrad $\cotetrad^{\mu}{}_{A}$ that are related with the metric of spacetime through
\begin{equation}
    g_{\mu\nu} = \eta_{AB} \tetrad^A{}_{\mu} \tetrad^B{}_{\nu}, \hspace{4mm}   \eta_{AB} = g_{\mu\nu} \cotetrad^{\mu}{}_A \cotetrad^{\nu}{}_{B}.
    \label{eq:fundrel}
\end{equation}
The tetrad and cotetrad components also satisfy orthonormality relations
\begin{equation}
\tetrad^A{}_{\mu}  \cotetrad^{\mu}{}_B = \delta^{A}_{B}, \hspace{4mm} \tetrad^A{}_{\mu}  \cotetrad^{\nu}{}_A = \delta^{\nu}_{\mu}.\label{eq.2}
\end{equation}

Lorentz indices can be transformed into  spacetime indices and vice versa by contraction with a tetrad or cotetrad components in the following way: a spacetime index ${}^\mu$ becomes a Lorentz index ${}^A$ through contraction with a tetrad $\theta^A{}_\mu$, while a spacetime index ${}_\mu$ becomes a Lorentz index ${}_A$ through contraction with an inverse tetrad $e_A{}^\mu$. Lorentz indices are raised and lowered with the Minkowski metric, while spacetime indices are raised and lowered with the spacetime metric.

In addition we introduce a curvatureless, metric compatible spin connection $\spinconnection^A{}_{B\mu }$, whose components are  defined as 
\begin{equation}
\label{eq:spinconnlor}
    \spinconnection^A{}_{B\mu} = -\left(\Lorentz^{-1}\right)^C{}_B \partial_\mu \Lorentz_C{}^A, 
\end{equation}
where $\Lorentz_C{}^A$ are matrices satisfying properties of Lorentz matrices. The spin connection enters the teleparallel action and, thus, in this formulation the Lorentz matrices are treated as dynamical fields \cite{Golovnev:2021omn} \footnote{Other covariant formulations are reviewed in \cite{Golovnev:2017dox}.}. 

The main building block used in teleparallel theories of gravity is the torsion tensor, which depends on both the tetrad and spin connection as
\begin{equation}
    T^{A}_{\ \ \mu\nu}=\partial_{\mu}\theta^{A}_{\ \ \nu}-\partial_{\nu}\theta^{A}_{\ \ \mu}+\omega^{A}_{\ \ B\mu}\theta^{B}_{\ \ \nu}-\omega^{A}_{\ \ B \nu}\theta^{B}_{\ \ \mu}.\label{eq.5}
\end{equation}
With it, we can build the torsion scalar 
\begin{align}
    \mathbb{T}=\frac{1}{4}T^\rho{}_{\mu\nu}T_\rho{}^{\mu\nu}+\frac{1}{2}T^\rho{}_{\mu\nu}T^{\nu\mu}{}_\rho -T^\rho{}_{\mu\rho}T^{\sigma\mu}{}_\sigma, \label{eq.6}
\end{align}
which is related with the Ricci scalar from GR by only a boundary term
\begin{equation}
\mathbb R = - \mathbb T + 2 \tetrad \partial_{\mu} (\cotetrad T^{\mu} ),\label{eq.7}
\end{equation}
with $T^{\mu} = T^{\alpha}{}_{\alpha}{}^{\mu}$, $\tetrad = \text{det}(\tetrad^{A}{}_{\mu} )$ as the determinant of the tetrad and $e$ the determinant of its inverse. 
The torsion scalar can alternatively be written as
\begin{align}
\mathbb T = T^{\rho\mu\nu}S_{\rho\mu\nu},\label{eq.8}
\end{align}
where the so-called superpotential is defined in terms of the torsion tensor as
\begin{align}
    S_{\rho\mu\nu}=\frac{1}{2}T_{\rho\mu\nu}+T_{[\mu\nu]\rho}+2g_{\rho[\mu}T^\sigma{}_{\nu]\sigma}.\label{eq.9}
\end{align} 
The torsion scalar defines the TEGR Lagrangian $\mathcal{L}_{\mathrm{TEGR}}=-\frac{1}{2\kappa}\tetrad \mathbb T$, where $\kappa=\frac{8\pi G}{c^{4}}$. If we take the $\mathbb R$ in terms of the torsion scalar and the boundary term in \eqref{eq.7}, and replace it into the Einstein-Hilbert action, then the result is the action for TEGR,
\begin{align}
    S_\mathrm{TEGR}=-\frac{1}{2\kappa}\int \mathrm{d}^4 x \tetrad \mathbb{T}.\label{eq.10}
\end{align}

Since the boundary term is integrated out, we obtain a gravitational theory with the same equations of motion as GR. The equations of motion are obtained varying with respect to the tetrad $\tetrad^{A}{}_{\nu}$, and they are given in vacuum by
\begin{align}
    \cotetrad^{\nu}_{A} \mathbb T - 2 e  \partial_{\lambda}( \tetrad \cotetrad^{\sigma}{}_{A} S_{\sigma}{}^{\lambda\nu} ) - 2 \cotetrad^{\mu}{}_{A} T^{\rho}{}_{\sigma\mu} S_{\rho}{}^{\sigma\nu} = 0. 
    \label{eq.11}
\end{align}
The torsion scalar \eqref{eq.6} is the building block for modified gravity theories. For instance, relaxing the coefficients in front of the three terms quadratic in the torsion tensor, gives the theory so-called ``new general relativity''. An even more popular theory considered in the literature is to take an arbitrary function of the torsion scalar. This theory is referred to as $f(\mathbb{T})$ gravity which is analogous to $f(\mathbb R)$ gravity, but with essentially different physical content.

We are interested in studying the dynamics of the equations of motion \eqref{eq.11} from the Hamiltonian point of view. In particular, we would like to perform a 3+1 decomposition of the equations of motion, which we will achieve by computing  Hamilton's equations. Although it is also possible to perform such split directly in the Lagrangian equations of motion \eqref{eq.11}, both methods give equivalent results, with the difference that with the Hamiltonian approach we get the chance to deepen into the structure of the Hamiltonian for TEGR, and preparing the formalism in order to be applied to modified teleparallel gravities previously mentioned.

\subsection{3+1 decomposition}

Our first step into getting Hamilton's equations of motion for TEGR consists in performing a proper 3+1 decomposition. This issue is more subtle than for metric-based theories, since our fundamental variable is now the tetrad field, which has more independent components. 
First we slice the four-dimensional manifold described by the metric $g_{\mu\nu}$ into three-dimensional hypersurfaces of constant time $\Sigma_t$ that are equipped with a three-dimensional induced metric $\inducedmetric_{ij}$ \footnote{Such decomposition assumes that the tetrad respect the conditions for a proper foliation. This issue will be addressed in a forthcoming paper \cite{BGSW}}. We also introduce the lapse $\alpha$ and shift $\beta^{i}$ functions, therefore the four-dimensional metric is then the usual ADM one:
\be
g_{00}=-\lapse^2+\shift^i\shift^j\inducedmetric_{ij}, \quad g_{0i}=\shift_i,\quad g_{ij}=\inducedmetric_{ij}
\label{eq:ADMmetric}
\ee
and the inverse metric
\be
g^{00}=-\frac{1}{\lapse^2},\quad g^{0i}=\frac{\shift^i}{\lapse^2},\quad g^{ij}=\inducedmetric^{ij}-\frac{\shift^i\shift^j}{\lapse^2}.
\label{eq:ADMinv}
\ee
We will consider the spatial components $\theta^{A}{}_{i}$ of the tetrad as canonical variables instead of the induced metric. However, the latter can be written in terms of the former by virtue of
\be
\tetrad^A{}_i\tetrad^B{}_j\eta_{AB}=\inducedmetric_{ij}.
\ee
A possible ADM decomposition of the temporal part of the tetrad can be written as
\be
\tetrad^A{}_0=\lapse\normalvector^A+\shift^i\tetrad^A{}_i.
\label{ADMtetrad}
\ee
Here we have introduced the vector $\normalvector^{A}$ with Lorentz indices. In order to recover the ADM metric from the tetrad \eqref{ADMtetrad}, this vector needs to satisfy the condition
\be
\eta_{AB}\normalvector^A\normalvector^B=\normalvector_A\normalvector^A=-1, \label{eq.16}
\ee
together with being orthonormal to the spatial part of the tetrad, that is,
\be 
\eta_{AB}\normalvector^B\tetrad^A{}_i=\normalvector_A\tetrad^A{}_i=0. \label{eq.17}
\ee
Notice that the vector $\normalvector^{A}$ that satisfies all these properties can be written as (see, for instance, \cite{Castellani:1981ue}),
\be 
\xi^{A} = -\dfrac16 \epsilon^{A}{}_{BCD} \tetrad^{B}{}_{i} \tetrad^{C}{}_{j} \tetrad^{D}{}_{k} \epsilon^{ijk}.
\ee 
Finally, the ADM split of the inverse tetrad $\cotetrad_{A}{}^{\mu}$ can be consistently proposed as
\be
\cotetrad_A{}^0=-\frac{1}{\alpha}\normalvector_A, \indent  \cotetrad_A{}^i=\tetrad_A{}^i+\normalvector_A\frac{\shift^i}{\lapse}.
\label{ADMinvE}
\ee
An important remark is that the rhs of the second equation in \eqref{ADMinvE} defines the object $\theta_{A}{}^{i}$, which is the shorthand notation for $\theta_{A}{}^{i} = \eta_{AB} \gamma^{ij} \tetrad^{B}{}_{j}$. In our main results we will try to make little use of the $\theta_{A}{}^{i}$, since it can be confused with the genuine inverse tetrad components $e_{A}{}^{i}$. Last but not least, an important and useful expression is 
\begin{align}
    \eta_{AD}\inducedmetric^{kl}\tetrad^C{}_k \tetrad^D{}_l = \tetrad^{C}{}_{k} \tetrad_{A}{}^{k} = \delta^C_A+\eta_{AD} \normalvector^C \normalvector^D.
\end{align}
From the ADM decomposition proposed we can observe that our canonical variables will be $(\alpha,\beta^{i},\tetrad^{A}{}_{i})$, spanning the 16 independent components of the tetrad field. Therefore, we can compute the 3+1 Lagrangian for TEGR. This is not a simple task, but after some efforts it is obtained the following result
\be
 \mathcal{L}_\mathrm{TEGR}=\frac{\sqrt{\inducedmetric}}{2\lapse}M^{i \ j}_{\ A \ B}T^A{}_{0i}T^B{}_{0j} -\frac{\sqrt{\inducedmetric}}{\lapse}T^A{}_{0i}T^B{}_{kl}\cdot \left[M^{i \ l}_{\ A \ B}\shift^k-\frac{\lapse}{\kappa}\inducedmetric^{il}\left(\frac{1}{2}\normalvector_B\tetrad_A{}^k-\normalvector_A\tetrad_B{}^k \right) \right] +L_S,
 \label{L3p1}
\ee
where the time derivatives of the tetrad field are encoded in the $T^{A}{}_{0i}$ components of the torsion tensor. The tensor $M^{i \ j}_{\ A \ B}$ accompanying the term quadratic in velocities is given by
\be
\begin{split}
    M^{i \ j}_{\ A \ B}=\frac{1}{\kappa}\left(\frac{1}{2}\inducedmetric^{ij}\eta_{AB}+\frac{1}{2}\normalvector_A\normalvector_B\inducedmetric^{ij}+\frac{1}{2}\tetrad_A{}^j\tetrad_B{}^i-\tetrad_A{}^i\tetrad_B{}^j\right),\label{eq.20}
\end{split}
\ee
while a term depending only on spatial derivatives of the tetrad can be written as
\be
\begin{split}
    L_S=\frac{\sqrt{\inducedmetric}}{\lapse}T^A{}_{ij}T^B{}_{kl}\shift^i\left[\frac{1}{2}M^{j \ l}_{\ A \ B}\shift^k-\frac{\lapse}{\kappa}\inducedmetric^{jl}\left(\frac{1}{2}\normalvector_B\tetrad_A{}^k-\normalvector_A\tetrad_B{}^k \right) \right] +\frac{\lapse\sqrt{\inducedmetric}}{2\kappa}{}^3\mathbb{T}.\label{eq.21}
\end{split}
\ee
Here we have defined the spatial part of the torsion scalar as ${}^3\mathbb{T}$
\be
\begin{split}
    {}^3\mathbb{T}&=H_{AB}{}^{ijkl}T^A{}_{ij}T^B{}_{kl}\\
    & =\left(-\frac{1}{4}\eta_{AB} \inducedmetric^{k[i}\inducedmetric^{j]l}+\frac{1}{2}\tetrad_B{}^{[i}\inducedmetric^{j][k}\tetrad_A{}^{l]}-\tetrad_A{}^{[i}\inducedmetric^{j] [k}\tetrad_B{}^{l]} \right)T^A{}_{ij}T^B{}_{kl}.\label{eq.22}
\end{split}
\ee
From this Lagrangian we must obtain the canonical momenta, which will correspond to the 16 phase space functions $(\pi, \pi_i, \pi_{A}{}^{i})$ associated to lapse, shift, and spatial part of the tetrad. The canonical momenta are easily obtained from here, since the only components containing time derivatives of the spatial tetrad $\partial_0 \tetrad^{A}{}_{i}$ are those containing $T^{A}{}_{0i}$. Therefore, they are computed from our previous Lagrangian as
\begin{equation}
\momenta_{A}{}^{i} = \dfrac{\partial L}{\partial_0  \tetrad^A{}_\mu} = \dfrac{\partial L}{\partial T^{A}{}_{0i} } = \dfrac{\sqrt{\gamma} }{ \alpha} \left[  M^{i \ l}_{\ A \ B} (T^{B}{}_{0l} - T^{B}{}_{ml}\shift^m) + \dfrac{\alpha}{\kappa}T^{B}{}_{ml} \inducedmetric^{il}\left(\dfrac12 \normalvector_{B}\tetrad_{A}{}^{m} - \normalvector_{A}\tetrad_{B}{}^{m} \right)  \right].
\label{piT}
\end{equation}
The conjugate momenta associated with lapse and shift form part of a primary constraints, since the Lagrangian does not present time derivatives of them, therefore
\begin{equation}
\label{Calph}
    {}^\alpha C={}^\alpha \pi:=\frac{\partial L}{\partial \partial_{0} \alpha}\approx 0,
    \end{equation}
    \begin{equation}
    \label{Cbeth}
     {}^\beta C_i={}^\beta \pi_i:=\frac{\partial L}{\partial \partial_0 \beta^i}\approx 0. 
\end{equation}
We will denote them collectively as $C_{A} = ({}^\alpha C, {}^\beta C_i )$, and they will have associated Lagrange multipliers denoted by $({}^{\alpha}\lambda, {}^{\beta}\lambda^i )$. In other works available in the literature, the choice of canonical variables is $(\tetrad^{A}{}_{0}, \tetrad^{A}{}_{i})$ instead of $(\lapse,\shift^{i},\tetrad^{A}{}_{i})$, therefore these trivial primary constraints turn out to be $\momenta_{A}{}^{0} \approx 0$ due to the nonappearance of time derivatives of $\tetrad^{A}{}_{0}$. These primary constraints generically appear not only in TEGR but in any teleparallel theory based on the torsion tensor/tetrad field.

In addition to these momenta, there must be considered the momenta associated to the Lorentz matrices introduced through the teleparallel spin connection \eqref{eq:spinconnlor}. In \cite{Golovnev:2021omn} the momenta $P^{B}{}_{A}$ have been taken as the variation with respect to $\partial_0 \Lambda^{A}{}_{B}$
\begin{align}
    P^B{}_A:=\frac{\partial L}{\partial_0 \Lambda^A{}_B}=\momenta_C{}^i \eta_{AD}\left(\Lambda^{-1} \right)_E{}^B\eta^{C[E}\tetrad^{D]}{}_i.
\end{align}
There is no assumption of antisymmetry on the matrices, but their combination with $P^{B}{}_{A}$ in a primary constraint restricts the free components of the Lorentz matrices. These primary constraints read
\begin{align}
\label{eq:CovConstraints}
C^{AB} = P^{[A}{}_D\eta^ {B]C}\Lorentz_C{}^D+\momenta_C{}^i\eta^{C[B}\tetrad^{A]}{}_i \approx 0.
\end{align}
It has been proved that these extra constraints have zero Poisson bracket with the remaining primary constraints, and since due to their simple form it is expected this to happen with the Hamiltonian too, it is an educated guess to postulate that they are first class, as shown in \cite{Golovnev:2021omn}.

With these tools we are able to tackle the computation of the Hamiltonian for TEGR in the next section, from which we will extract Hamilton's equations. Note that the closed form of the covariant Hamiltonian in teleparallel theories has only been presented in the PhD thesis \cite{thesis}.

\section{Hamiltonian for TEGR}
\label{sec:Ham}

In order to obtain the Hamiltonian for TEGR from the $3+1$ Lagrangian \eqref{L3p1}, we must solve the velocities in terms of the momenta in \eqref{piT}. For this, it is useful to make a decomposition of the velocities and conjugate momenta into irreducible parts under the rotation group $\mathcal{O}(3)$ \cite{Blixt19,Blixt:2020ekl}. Such decomposition reads
\begin{align}
    \dot{\tetrad}^A{}_i={}^{\mathcal{V}}{}\dot{\tetrad}_i\normalvector^A+{}^{\mathcal{A}}{} \dot{\tetrad}_{ji}\inducedmetric^{kj}\tetrad^A{}_k+{}^{\mathcal{S}}{} \dot{\tetrad}_{ji}\inducedmetric^{kj}\tetrad^A{}_k+{}^{\mathcal{T}}{} \dot{\tetrad}\tetrad^A{}_i,
    \label{eq:velocVAST}
\end{align}
\begin{align}
    \momenta_A{}^i={}^{\mathcal{V}}{}\momenta^i\normalvector_A+{}^{\mathcal{A}}{} \momenta^{ji}\tetrad^B{}_j\eta_{AB}+{}^{\mathcal{S}}{} \momenta^{ji}\tetrad^B{}_j \eta_{AB}+{}^{\mathcal{T}}{} \momenta\tetrad^B{}_j \eta_{AB}\inducedmetric^{ij}.
    \label{eq:momVAST}
\end{align}

We can also write the variables of the irreducible decomposition in terms of the original canonical variables $\dot{\tetrad}^{A}{}_i$ and $\pi_{A}{}^{i}$ as
\begin{align}     
\begin{split}     
     {}^{\mathcal{S}}{} \dot{\tetrad}_{ji}&=\dot{\tetrad}_{(ji)}- \frac{1}{3}\dot{\tetrad}^A{}_k\tetrad^B{}_l\eta_{AB}\inducedmetric^{kl}\inducedmetric_{ij}=\frac{1}{2} \dot{\tetrad}^A{}_i\tetrad^B{}_j\eta_{AB} +\frac{1}{2}\dot{\tetrad}^A{}_j\tetrad^B{}_i\eta_{AB} -\frac{1}{3}\dot{\tetrad}^A{}_k\tetrad^B{}_l\eta_{AB}\inducedmetric^{kl}\inducedmetric_{ij}, \\
      {}^{\mathcal{T}}{} \dot{\tetrad}&=\frac{1}{3} \dot{\tetrad}^A{}_i \tetrad^B{}_j\eta_{AB}\inducedmetric^{ij},\\
   {}^{\mathcal{V}}{} \dot{\tetrad}_i & =-\normalvector_A \dot{\tetrad}^A{}_i,\\
       {}^{\mathcal{A}}{} \dot{\tetrad}_{ji} & = \dot{\tetrad}_{[ji]}=\frac{1}{2} \dot{\tetrad}^A{}_i \tetrad^B{}_j\eta_{AB} -\frac{1}{2}\dot{\tetrad}^A{}_j \tetrad^B{}_i\eta_{AB},
       \label{eq.27}
  \end{split} \end{align}
and
\begin{equation}
    \begin{split}
        {}^{\mathcal{S}}\momenta^{ji}&=\momenta^{(ji)}-\frac{1}{3}\momenta_A{}^k\tetrad^A{}_k\inducedmetric^{ij}=\frac{1}{2}\momenta_A{}^i\tetrad^A{}_k\inducedmetric^{jk}+\frac{1}{2}\momenta_A{}^j\tetrad^A{}_k\inducedmetric^{ik}-\frac{1}{3}\momenta_A{}^k\tetrad^A{}_k\inducedmetric^{ij}, \\
        {}^{\mathcal{T}} \momenta&=\frac{1}{3} \momenta_A{}^i \tetrad^A{}_i,\\
        {}^{\mathcal{V}} \momenta^i &=-\normalvector^A \momenta_A{}^i,\\
        {}^{\mathcal{A}} \momenta^{ji} &=\momenta^{[ji]}=\frac{1}{2}\momenta_A{}^i \tetrad^A{}_k\inducedmetric^{jk}-\frac{1}{2}\momenta_A{}^j \tetrad^A{}_k\inducedmetric^{ik}.
    \end{split}
        \label{symtrcMomenta}
\end{equation}

With this irreducible decomposition at hand, the TEGR primary constraints are  obtained from \eqref{piT} as
\begin{align} 
  {}^{\mathcal{V}}C^i =  \frac{{}^{\mathcal{V}} \pi^i\kappa}{\sqrt{\inducedmetric}}+T^B{}_{jk}\inducedmetric^{ik}\inducedmetric^{jl}\tetrad^A{}_l\eta_{AB}\approx 0, \label{Cvek}
\end{align}
\begin{align}
   {}^{\mathcal{A}}C^{ij} =  \frac{{}^{\mathcal{A}} \pi^{ij}\kappa}{\sqrt{\inducedmetric}}-\dfrac12 \inducedmetric^{ik}\inducedmetric^{jl}T^B{}_{kl}\normalvector_B\approx 0.\label{Casy}
\end{align}
In order to write the canonical Hamiltonian density, we must invert the velocities $T^{A}{}_{0i}$ in \eqref{piT} as a function of the momenta $\momenta_{A}{}^{i}$. This is facilitated by the irreducible decomposition that was introduced (details can be seen in \cite{Blagojevic00,Blixt19}). We find the Moore-Penrose pseudoinverse of M which reads
\begin{align}
    \begin{split}
        \left(M^{-1} \right)^{\ A \ C}_{ i \ k}=\frac{\kappa}{2}\left(\inducedmetric_{ik}\inducedmetric^{mn}\tetrad^A{}_m \tetrad^C{}_n+\tetrad^A{}_k\tetrad^C{}_i-\tetrad^A{}_i \tetrad^C{}_k\right).
    \end{split}
\end{align}
With this \eqref{piT} can be inverted as
\begin{align}
    \begin{split}
        T^C{}_{0k}=\left(M^{-1} \right)^{\ A \ C}_{i \ k}\frac{\alpha}{\sqrt{\inducedmetric}}\pi_A{}^i+T^C{}_{mk}\shift^m-\frac{\lapse}{\kappa}\left(M^{-1} \right)^{\ A \ C}_{i \ k}T^B{}_{ml}\inducedmetric^{il}\left(\frac{1}{2}\normalvector_B \tetrad_A{}^m -\normalvector_A \tetrad_B{}^m \right),
    \end{split}
\end{align}
so
\begin{align}
    \begin{split}
        \dot{\tetrad}^C{}_{k}-\left(\Lambda^{-1} \right)^A{}_B \dot{\Lambda}_A{}^C\tetrad^B{}_k&=\partial_k\tetrad^C{}_0+\spinconnection^C{}_{Dk}\tetrad^D{}_0+\left(M^{-1} \right)^{\ A \ C}_{i \ k}\frac{\alpha}{\sqrt{\inducedmetric}}\pi_A{}^i+T^C{}_{mk}\shift^m\\
        &-\frac{\lapse}{\kappa}\left(M^{-1} \right)^{\ A \ C}_{i \ k}T^B{}_{ml}\inducedmetric^{il}\left(\frac{1}{2}\normalvector_B \tetrad_A{}^m -\normalvector_A \tetrad_B{}^m \right).
    \end{split}
\end{align}
Note that due to the primary constraints \eqref{eq:CovConstraints}, the tetrad velocities and momenta velocities need to be inverted together. The Hamiltonian density is thus given by
\begin{align}
    \mathcal{H}_c=  \momenta_A{}^i\left(\dot{\tetrad}^A{}_{k}-\left(\Lambda^{-1} \right)^C{}_B \dot{\Lambda}_C{}^A\tetrad^B{}_i\right)-\mathcal{L},
\end{align}
and the primary Hamiltonian density is obtained from the canonical Hamiltonian by adding a linear combination of primary constraints $C_{a}$ multiplied by Lagrange multipliers $\lambda^{a}$
\begin{align}
    \mathcal{H}_\mathrm{TEGR}= \mathcal{H}_c -\lambda^{a}C_{a} =  \momenta_A{}^i\left(\dot{\tetrad}^A{}_{k}-\left(\Lambda^{-1} \right)^C{}_B \dot{\Lambda}_C{}^A\tetrad^B{}_i\right)-\mathcal{L} - \lambda^{a} C_{a}.
\end{align}
Note that the (canonical, primary, etc) Hamiltonian is obtained by integrating the (canonical, primary, etc) Hamiltonian density over space, that is $H=\int d^3x \mathcal{H}$. We have abbreviated the set of primary constraints into the array $C_{a} = (C_A, C_{AB}, {}^{\mathcal{V}}C^i , {}^{\mathcal{A}}C^{ij})$, where all the components are given respectively by Eqs.\eqref{Calph},\eqref{Cbeth},\eqref{eq:CovConstraints},\eqref{Cvek}, and \eqref{Casy}. These are all our primary constraints that must be included in the primary Hamiltonian with arbitrary Lagrange multipliers $\lambda^{a} = (\lambda^{A}, \lambda^{AB}, {}^{\mathcal{V}}\lambda_i , {}^{\mathcal{A}}\lambda_{ij})$.

After some computations, we obtain the primary Hamiltonian for covariant TEGR
\begin{align}
\begin{split}
    \mathcal{H}_\mathrm{TEGR}=\lapse  \left[ \frac{\kappa}{2\sqrt{\inducedmetric}} {}^{\mathcal{S}}{\momenta}_{ij} {}^{\mathcal{S}} \momenta^{ij} -\frac{3\kappa}{4\sqrt{\inducedmetric}}{}^{\mathcal{T}} \momenta {}^{\mathcal{T}} \momenta -\frac{\sqrt{\inducedmetric}}{2\kappa}{}^3 \mathbb{T}-\normalvector^A\partial_i \momenta_A{}^i+\momenta_A{}^i\spinconnection^A{}_{Bi}\normalvector^B \right] \\
    +\shift^j\left[ -\tetrad^A{}_j \partial_i \momenta_A{}^i+\momenta_A{}^i\spinconnection^A{}_{Ci} \tetrad^C{}_j-\momenta_A{}^i T^A{}_{ij} \right] -\lambda^{A}C_{A} -\lambda_{AB}\left( P^{[A}{}_D\eta^ {B]C}\Lorentz_C{}^D+\momenta_C{}^i\eta^{C[B}\tetrad^{A]}{}_i \right) \\
  -{}^{\mathcal{V}}\lambda_i\left(\frac{{}^{\mathcal{V}} \pi^i\kappa}{\sqrt{\inducedmetric}}+T^B{}_{jk}\inducedmetric^{ik}\inducedmetric^{jl}\tetrad^A{}_l\eta_{AB}\right)-{}^{\mathcal{A}}\lambda_{ij}\left(\frac{{}^{\mathcal{A}} \pi^{ij}\kappa}{\sqrt{\inducedmetric}}-\dfrac12 \inducedmetric^{ik}\inducedmetric^{jl}T^B{}_{kl}\normalvector_B \right)+\partial_i \left(\momenta_A{}^i\tetrad^A{}_0 \right).
  \label{eq.31}
\end{split}
\end{align}
The boundary term contains nonlinearities in lapse and shift, and thus it will be dropped for the rest of the article:
\begin{align}
\begin{split}
    \mathcal{H}_\mathrm{TEGR}=\lapse  \left[ \frac{\kappa}{2\sqrt{\inducedmetric}} {}^{\mathcal{S}}{\momenta}_{ij} {}^{\mathcal{S}} \momenta^{ij} -\frac{3\kappa}{4\sqrt{\inducedmetric}}{}^{\mathcal{T}} \momenta {}^{\mathcal{T}} \momenta -\frac{\sqrt{\inducedmetric}}{2\kappa}{}^3 \mathbb{T}-\normalvector^A\partial_i \momenta_A{}^i+\momenta_A{}^i\spinconnection^A{}_{Bi}\normalvector^B \right] \\
    +\shift^j\left[ -\tetrad^A{}_j \partial_i \momenta_A{}^i+\momenta_A{}^i\spinconnection^A{}_{Ci} \tetrad^C{}_j-\momenta_A{}^i T^A{}_{ij} \right] -\lambda^{A}C_{A} -\lambda_{AB}\left( P^{[A}{}_D\eta^ {B]C}\Lorentz_C{}^D+\momenta_C{}^i\eta^{C[B}\tetrad^{A]}{}_i \right) \\
  -{}^{\mathcal{V}}\lambda_i\left(\frac{{}^{\mathcal{V}} \pi^i\kappa}{\sqrt{\inducedmetric}}+T^B{}_{jk}\inducedmetric^{ik}\inducedmetric^{jl}\tetrad^A{}_l\eta_{AB} \right)-{}^{\mathcal{A}}\lambda_{ij}\left(\frac{{}^{\mathcal{A}} \pi^{ij}\kappa}{\sqrt{\inducedmetric}}-\dfrac12 \inducedmetric^{ik}\inducedmetric^{jl}T^B{}_{kl}\normalvector_B \right).
  \label{eq.32}
\end{split}
\end{align}
In the Weitzenb\"{o}ck gauge the expression for the Hamiltonian reduces to 
\begin{align}
\begin{split}
    \mathcal{H}_\mathrm{TEGR}=\lapse  \left[ \frac{\kappa}{2\sqrt{\inducedmetric}} {}^{\mathcal{S}}{\momenta}_{ij} {}^{\mathcal{S}} \momenta^{ij} -\frac{3\kappa}{4\sqrt{\inducedmetric}}{}^{\mathcal{T}} \momenta {}^{\mathcal{T}} \momenta -\frac{\sqrt{\inducedmetric}}{2\kappa}{}^3 \mathbb{T}-\normalvector^A\partial_i \momenta_A{}^i \right] +\shift^j\left[ -\tetrad^A{}_j \partial_i \momenta_A{}^i-\momenta_A{}^i T^A{}_{ij} \right] \\ -\lambda^{A}C_{A}  -{}^{\mathcal{V}}\lambda_i\left(\frac{{}^{\mathcal{V}} \pi^i\kappa}{\sqrt{\inducedmetric}}+T^B{}_{jk}\inducedmetric^{ik}\inducedmetric^{jl}\tetrad^A{}_l\eta_{AB} \right)-{}^{\mathcal{A}}\lambda_{ij}\left(\frac{{}^{\mathcal{A}} \pi^{ij}\kappa}{\sqrt{\inducedmetric}}-\dfrac12 \inducedmetric^{ik}\inducedmetric^{jl}T^B{}_{kl}\normalvector_B  \right).
    \label{eq.33}
\end{split}
\end{align}
In order to derive Hamilton's equation in terms of the original variables (not the irreducible decomposition ones), we will go back to the canonical momenta $\momenta_A{}^i$ by inverting the Eq. \eqref{symtrcMomenta}, obtaining
\be 
2 {}^{\mathcal{S}}{\momenta}_{ij} {}^{\mathcal{S}} \momenta^{ij}-3 {}^{\mathcal{T}} \momenta {}^{\mathcal{T}} \momenta = \momenta_A{}^i\momenta_B{}^l\tetrad^A{}_k \tetrad^B{}_j\inducedmetric^{jk}\inducedmetric_{li} +\momenta_A{}^i\momenta_B{}^j\tetrad^A{}_j\tetrad^B{}_i-\momenta_A{}^i\momenta_B{}^j\tetrad^A{}_i\tetrad^B{}_j .\label{eq.34}
\ee 
The primary Hamiltonian can thus be explicitly written in terms of the conjugate momenta  $\momenta_A{}^i$ of the tetrad in the following way:
\begin{align}
\begin{split}
    \mathcal{H}_\mathrm{TEGR}&=\lapse  \left[\frac{ \kappa}{4\sqrt{\inducedmetric}}\left[\momenta_A{}^i\momenta_B{}^l\tetrad^A{}_k \tetrad^B{}_j\inducedmetric^{jk}\inducedmetric_{li}  +\momenta_A{}^i\momenta_B{}^j\tetrad^A{}_j\tetrad^B{}_i -\momenta_A{}^i\momenta_B{}^j\tetrad^A{}_i\tetrad^B{}_j \right]  \right. \\
    & \left. -\frac{\sqrt{\inducedmetric}}{2\kappa}{}^3 \mathbb{T}-\normalvector^A\partial_i \momenta_A{}^i \right] +\shift^j\left[ -\tetrad^A{}_j \partial_i \momenta_A{}^i-\momenta_A{}^i T^A{}_{ij} \right] -\lambda^{A}C_{A} \\
    &-{}^{\mathcal{V}} \lambda_i\left[ -\dfrac{\kappa}{\sqrt{\inducedmetric}} \normalvector^A \momenta_A{}^i +T^B{}_{jk}\inducedmetric^{ik}\inducedmetric^{jl}\tetrad^A{}_l\eta_{AB} \right]\\
    &-{}^{\mathcal{A}} \lambda_{ij}\left[\dfrac{\kappa}{2\sqrt{\inducedmetric}}\tetrad^A{}_k(\momenta_A{}^j\inducedmetric^{ik}-\momenta_A{}^i\inducedmetric^{jk})  -\dfrac12 \inducedmetric^{ik}\inducedmetric^{jl}T^B{}_{kl}\normalvector_B \right].
    \label{eq.35}
\end{split}
\end{align}
\subsection{Comparing standard GR and TEGR canonical variables}
In Appendix \ref{app:B} the standard way to derive the Hamiltonian and Hamilton's equations in GR (with the metric formulation of the EH action) is presented. It is evident from \eqref{eq:velocVAST} and \eqref{eq:momVAST} that there is not a one to one relationship among the canonical variables of standard GR and TEGR. From these equations and \eqref{eq:fundrel}, we can attempt to derive a relation among the velocities $\dot{\tetrad}^{A}{}_i$ and $\dot{\gamma}_{ij}$ as
\begin{equation}
    \dot{\inducedmetric}_{ij} = \eta_{AB}( \dot{\tetrad}^{A}{}_i \tetrad^{B}{}_j + \tetrad^{A}{}_i \dot{\tetrad}^{B}{}_j)
    \label{eq.36}
\end{equation}

Using \eqref{eq:velocVAST} here, we realize that the time derivative of the induced metric depends only on the irreducible parts that are symmetric, that is
\begin{equation}
\dot{\gamma}_{ij}= 2({}^{\mathcal{S}}{}\dot{\tetrad}_{ij}+{}^{\mathcal{T}}{}\dot{\tetrad}\gamma_{ij}) \label{eq.37}
\end{equation}
From the definition of the canonical momenta for the tetrad $\pi_{A}{}^{i}$, we can then explicitly write its relation with the GR momenta $\pi^{ij}$ as
\begin{equation}
\pi_A{}^{i} = \pi^{ij}\tetrad^{B}{}_j \eta_{AB} + {}^{\mathcal{V}}{}\momenta^{i}\normalvector_A + {}^{\mathcal{A}}{}\pi^{ji}\tetrad^{B}{}_{j}\eta_{AB} \label{eq.38}
\end{equation}
since the momenta for standard GR and TEGR are
\be 
\pi^{ij} = \dfrac{\partial L}{\partial \dot{\inducedmetric}_{ij} }~~, \ \ \momenta_A{}^{i} = \dfrac{\partial L}{\partial \dot{\tetrad}^{A}{}_{i}}.\label{eq.39}
\ee 

{}
\section{Hamilton's equations for TEGR}
\label{sec:HeqsW}
In this section we compute Hamilton's equations for TEGR in the Weitzenb\"{o}ck gauge (zero spin connection). The difference between the derivation presented here and the derivation of Hamilton's equations for GR is in the enlarged set of canonical variables. Of course, the Hamiltonian for TEGR differs in many aspects from the Hamiltonian for standard GR, since the boundary term modifies the canonical structure, but not the degrees of freedom \footnote{For more general discussion on the effect of boundary terms in the Hamiltonian formalism, see examples in \cite{Ferraro:2020tqk} and \cite{Golovnev:2021omn}.}. First, the phase space will be determined by the pairs of canonical variables $(\tetrad^{A}{}_{i}, \momenta_{B}{}^{j})$, therefore the fundamental Poisson brackets of two functions in the phase space $F(x)$ and $G(y)$ are defined as
\begin{equation}
    \{ F(x), G(y) \} = \int dz \left( \dfrac{\delta F(x) }{\delta \theta^{C}{}_{k}(z)} \dfrac{\delta G(y) }{\delta \pi_{C}{}^{k}(z) } - \dfrac{\delta F(x) }{\delta \pi_{C}{}^{k}(z) } \dfrac{\delta G(y) }{\delta \theta^{C}{}_{k}(z) } \right).\label{eq.40}
\end{equation}
Given a Hamiltonian $H(\tetrad^{A}{}_{i}, \momenta_{A}{}^{i}) = \int d^{3}x \mathcal{H}$ in the phase space, Hamilton's equations are
\begin{equation}
\dot{\tetrad}^{A}{}_{i} = \{ \tetrad^{A}{}_{i} , H \} = \dfrac{\delta H }{\delta \momenta_{A}{}^{i} }, \hspace{9mm}  \dot{\momenta}_{A}{}^{i} = \{ \momenta_{A}{}^{i} , H \} = - \dfrac{\delta H }{\delta  \tetrad^{A}{}_{i} }.
\label{eq.41}
\end{equation}
Analogously to GR, we obtain additional equations of motion from considering the dynamics of $(\alpha,\beta^{i})$ and their canonical momenta ${}^{\alpha}\pi$ and ${}^{\beta_i}\pi_i$, which have not been considered as Lagrange multipliers from the beginning, but belonging to the set of canonical variables. This is in contrast with the Lagrangian multipliers ${}^{\mathcal{V}}\lambda_i$ and ${}^{\mathcal{A}}\lambda_{ij}$ that accompany the extra primary constraints of TEGR (in comparison to GR), and do not have a momenta associated. Therefore, we will compute the variations of the Hamiltonian with respect to the canonical fields and their associated momenta, which is equivalent to the computation of the corresponding Poisson brackets of them with the Hamiltonian.

If we differentiate $H$ with respect to lapse we get
\begin{align}
    \begin{split}
        -{}^\alpha\dot{\pi}=\frac{\delta H}{\delta \alpha}=  \frac{\kappa}{2\sqrt{\inducedmetric}} {}^{\mathcal{S}}{\momenta}_{ij} {}^{\mathcal{S}} \momenta^{ij} -\frac{3\kappa}{4\sqrt{\inducedmetric}}{}^{\mathcal{T}} \momenta {}^{\mathcal{T}} \momenta -\frac{\sqrt{\inducedmetric}}{2\kappa}{}^3 \mathbb{T}-\normalvector^A\partial_i \momenta_A{}^i,
        \label{eq.42}
        \end{split}
\end{align}
where the rhs corresponds to the Hamiltonian constraint, also appearing in GR. Again we differentiate $H$ with respect to shift, obtaining
\begin{align}
 -{}^\beta\dot{\pi}_j=\frac{\delta H}{\delta \beta ^{j}}&=  -\tetrad^A{}_j \partial_i \momenta_A{}^i-\momenta_A{}^i T^A{}_{ij}, \label{eq.43}
\end{align}
which gives in the rhs the momenta constraint. 

Differentiating with respect to the spatial tetrads yields
\begin{align}
    \begin{split}
        -\dot{\pi}_A{}^i= \dfrac{\delta H}{\delta \tetrad^{A}{}_{i} } &= \dfrac{\lapse \kappa}{2\sqrt{\inducedmetric}}\left(  \momenta_A{}^j \momenta_B{}^i \tetrad^B{}_j-\momenta_A{}^i \momenta_B{}^j \tetrad^B{}_j+ \inducedmetric^{il}\inducedmetric_{jk}\momenta_A{}^j\momenta_B{}^k \tetrad^B{}_l\right)
        -\shift^i \partial_j \momenta_A{}^j\\
        & + {}^{\mathcal{V}}\lambda^l T^B{}_{kl}(\tetrad_{A}{}^k\tetrad_{B}{}^{i}+\normalvector_{A}\normalvector_{B}\gamma^{ki} )+\lapse \normalvector_A \inducedmetric^{ik}\tetrad^B{}_k\partial_j \momenta_B{}^j-{}^{\mathcal{V}}\lambda_j\frac{\kappa}{\sqrt{\inducedmetric}}\normalvector_A \inducedmetric^{ik}\tetrad^B{}_k \momenta_B{}^j \\
        &-\dfrac12{}^{\mathcal{A}}\lambda^{lk}T^B{}_{kl}\normalvector_A \tetrad_B{}^i+\frac{\kappa}{\sqrt{\inducedmetric}}{}^{\mathcal{A}}\lambda_{[jk]}\momenta_A{}^j\inducedmetric^{ik}\\
        &+\frac{\lapse \tetrad_A{}^i}{2  }\left(-\frac{\sqrt{\inducedmetric}}{\kappa}{}^3 \mathbb{T}+\frac{\kappa}{2\sqrt{\inducedmetric}}\momenta_B{}^j\momenta_D{}^k\left(\tetrad^B{}_j\tetrad^D{}_k-\tetrad^B{}_k\tetrad^D{}_j-\inducedmetric_{jk}\inducedmetric^{ln}\tetrad^B{}_l\tetrad^D{}_n \right) \right)\\
        &-\frac{\kappa}{\sqrt{\inducedmetric}} {}^{\mathcal{V}}\lambda_j\tetrad_A{}^i\normalvector^B\momenta_B{}^j +{}^{\mathcal{V}}\lambda_j T^{B}{}_{kl}\tetrad_B{}^{k}(\tetrad_A{}^{j}\inducedmetric^{il} + \tetrad_{A}{}^{l} \inducedmetric^{ij} ) +\frac{\kappa}{\sqrt{\inducedmetric}} {}^{\mathcal{A}}\lambda_{[lj]} \inducedmetric^{kl}\tetrad_A{}^i\tetrad^B{}_k\momenta_B{}^j \\
        &-\dfrac12{}^{\mathcal{A}}\lambda_{nj}T^B{}_{kl} \normalvector_B \left[\inducedmetric^{jl}(\tetrad_{A}{}^{n}\inducedmetric^{ik} + \tetrad_{A}{}^{k}\inducedmetric^{ni} ) + \inducedmetric^{nk}(\tetrad_{A}{}^{j} \inducedmetric^{il} + \tetrad_{A}{}^{l}\inducedmetric^{ji} ) \right]  \\
        &+\frac{\kappa}{\sqrt{\inducedmetric}}{}^{\mathcal{A}}\lambda_{[lj]} \pi_B{}^j \tetrad^B{}_k (\tetrad_{A}{}^{l}\inducedmetric^{ik} + \tetrad_{A}{}^{k}\inducedmetric^{li} ) \\
        & + \frac{\kappa \lapse}{2\sqrt{\inducedmetric}}\eta_{AC}\left(\inducedmetric^{kl}\tetrad^B{}_l\tetrad^D{}_k\tetrad^C{}_j \momenta_B{}^i\momenta_D{}^j -\inducedmetric_{jk}\inducedmetric^{lm}\inducedmetric^{in}\tetrad^B{}_n\tetrad^D{}_l\tetrad^C{}_m\momenta_B{}^j\momenta_D{}^k \right) \\
        &-2\partial_{l}\left(\shift^{[l}\momenta_A{}^{i]}-{}^{\mathcal{V}}\lambda^{[i} \tetrad_A{}^{l]}+\dfrac12 {}^{\mathcal{A}}\lambda^{[il]}\normalvector_A -\frac{\lapse\sqrt{\inducedmetric}}{\kappa}H_{CA}{}^{[mn][il]}T^C{}_{mn}\right) \\
        &-\frac{\lapse \sqrt{\inducedmetric}}{\kappa}T^B{}_{kl}T^C{}_{mn}\left( \tetrad_{A}{}^{m}H_{CB}{}^{inkl} + \tetrad_{A}{}^{n}H_{CB}{}^{mikl} + \tetrad_{A}{}^{k}H_{CB}{}^{mnil} + \tetrad_{A}{}^{l}H_{CB}{}^{mnki} \right.\\
        &\left. + \normalvector_C \normalvector_A \inducedmetric^{i[m}\inducedmetric^{n][k}\tetrad_{B}{}^{l]}  + \normalvector_B \normalvector_A \tetrad_{C}{}^{[m}\inducedmetric^{n][k}\inducedmetric^{l]i} \right)\\
        \label{eq.44}
    \end{split}
\end{align}

Now we differentiate with respect to conjugate momenta and we get 
\begin{align}
    \dot{\alpha}=\frac{\delta H}{\delta {}^\alpha \pi}=- {}^{\alpha}\lambda,
    \label{eq.47}
\end{align}
\begin{align}
    \dot{\beta}_i=\frac{\delta H}{\delta {}^\beta \pi^i}= - {}^{\beta}\lambda^{i}. \label{eq.48}
\end{align}
The previous two Hamilton equations indicate that the lapse and shift have an arbitrary evolution, since their time derivative gives an arbitrary Lagrange multiplier. The variation with respect to the canonical momenta of the tetrad gives
\begin{align}
    \begin{split}
        \dot{\theta}^A{}_i &= \dfrac{\delta H}{\delta \momenta_{A}{}^{i}}= \alpha\left[ \dfrac{\kappa}{2\sqrt{\gamma}}(2\momenta_{B}{}^{j}\tetrad^{A}{}_{[j}\tetrad^{B}{}_{i]} + \momenta_{B}{}^{j}\tetrad^{A}{}_{k}\tetrad^{B}{}_{l}\inducedmetric_{ij}\inducedmetric^{kl} ) +\partial_{i}\normalvector^{A} \right]\\
        & +\partial_i\left(\beta^{j}\tetrad^{A}_{j}\right)- \beta^{j}T^{A}{}_{ij} + {}^{\mathcal{V}} \lambda_i\dfrac{\kappa\normalvector^{A}}{\sqrt{\gamma} }   + {}^{\mathcal{A}}  \lambda_{[ij]}\dfrac{\kappa \inducedmetric^{jk}\tetrad^{A}{}_{k} }{\sqrt{\gamma} },
        \label{eq.49}
    \end{split}
\end{align}
which is the time evolution of the tetrad field, and it depends linearly in the canonical momenta. An analogous behavior occurs in GR for the time evolution of the induced metric in \eqref{gammatimeev}, which depends linearly on the corresponding canonical momenta.
Given a set of initial data, these equations have to be complemented by enforcing the initial data to satisfy the primary constraints \eqref{Calph},\eqref{Cbeth},\eqref{Cvek}, and \eqref{Casy}. Depending on the point of view, these can be considered as belonging to the set of equations of motion (Hamilton's equations) \cite{Golovnev:2022rui}.

\section{Hamilton's equation for the covariant TEGR}
\label{sec:Heqscov} 

We present the TEGR Hamiltonian in terms of the momenta of the vector, antisymmetric, symmetric trace free and trace (VAST) decomposition,
\begin{align}
\begin{split}
    H&=\lapse  \left[ \frac{\kappa}{2\sqrt{\inducedmetric}} {}^{\mathcal{S}}{\momenta}_{ij} {}^{\mathcal{S}} \momenta^{ij} -\frac{3\kappa}{4\sqrt{\inducedmetric}}{}^{\mathcal{T}} \momenta {}^{\mathcal{T}} \momenta -\frac{\sqrt{\inducedmetric}}{2\kappa}{}^3 \mathbb{T}-\normalvector^A\partial_i \momenta_A{}^i+\momenta_A{}^i\spinconnection^A{}_{Bi}\normalvector^B \right] \\
    &+\shift^j\left[ -\tetrad^A{}_j \partial_i \momenta_A{}^i+\momenta_A{}^i\spinconnection^A{}_{Ci} \tetrad^C{}_j-\momenta_A{}^i T^A{}_{ij} \right]-\lambda_{AB}\left( P^{[A}{}_D\eta^ {B]C}\Lorentz_C{}^D+\momenta_C{}^i\eta^{C[B}\tetrad^{A]}{}_i \right) \\
    &-{}^{\mathcal{V}}\lambda_i\left(\frac{{}^{\mathcal{V}}\pi^i\kappa}{\sqrt{\inducedmetric}}+T^B{}_{jk}\inducedmetric^{ik}\tetrad_B{}^j \right)-{}^{\mathcal{A}}\lambda_{ij}\left(\frac{{}^{\mathcal{A}}\pi^{ij}\kappa}{\sqrt{\inducedmetric}}-\dfrac12\inducedmetric^{il}\inducedmetric^{jk}T^A{}_{kl}\normalvector_A \right)-\lambda_{A}C^{A},
    \label{eq.52}
\end{split}
\end{align}
and the TEGR Hamiltonian back to the canonical variables
\begin{align}
\begin{split}
    H&=\lapse  \left[\frac{ \kappa}{4\sqrt{\inducedmetric}}\left[\momenta_A{}^i\momenta_B{}^l\tetrad^A{}_k \tetrad^B{}_j\inducedmetric^{jk}\inducedmetric_{li}  +\momenta_A{}^i\momenta_B{}^j\tetrad^A{}_j\tetrad^B{}_i-\momenta_A{}^i\momenta_B{}^j\tetrad^A{}_i\tetrad^B{}_j \right]  -\frac{\sqrt{\inducedmetric}}{2\kappa}{}^3 \mathbb{T}-\normalvector^A\partial_i \momenta_A{}^i \right. \\
    & \left. +\momenta_A{}^i\spinconnection^A{}_{Bi}\normalvector^B\right]+\shift^j\left[ -\tetrad^A{}_j \partial_i \momenta_A{}^i+\momenta_A{}^i\spinconnection^A{}_{Ci} \tetrad^C{}_j-\momenta_A{}^i T^A{}_{ij} \right]\\
    &-\lambda_{AB}\left( P^{[A}{}_D\eta^ {B]C}\Lorentz_C{}^D+\momenta_C{}^i\eta^{C[B}\tetrad^{A]}{}_i \right)-{}^{\mathcal{V}} \lambda_i\left[ -\frac{\kappa}{\sqrt{\inducedmetric}}\normalvector^A \momenta_A{}^i+T^B{}_{jk}\inducedmetric^{ik}\tetrad_B{}^j \right]\\
    &-{}^{\mathcal{A}} \lambda_{ij}\left[\frac{\kappa}{2\sqrt{\inducedmetric}}\tetrad^A{}_k(\momenta_A{}^j\inducedmetric^{ik} -\momenta_A{}^i\inducedmetric^{jk})-\dfrac12 \inducedmetric^{ik}\inducedmetric^{jl}T^B{}_{kl}\normalvector_B \right]-\lambda_{A}C^{A}.
    \label{eq.53}
\end{split}
\end{align}
Although the difference between the covariant Hamiltonian and the Weitzenb\"{o}ck-like one is complex, and it cannot be easily written as isolated terms in the spin connection, the variations of the main fields have a simple form. Let us take as an example the torsion tensor, 
\be
\begin{split}
T^A{}_{ij}&=\partial_i \tetrad^A{}_j-\partial_j\tetrad^A{}_i+\spinconnection^A{}_{Bi}\tetrad^B{}_j-\spinconnection^A{}_{Bj}\tetrad^B{}_i \\
&=\partial_i \tetrad^A{}_j-\partial_j\tetrad^A{}_i+\left(\Lorentz^{-1}\right)^C{}_B \partial_i \Lorentz_C{}^A\tetrad^B{}_j-\left(\Lorentz^{-1}\right)^C{}_B \partial_j \Lorentz_C{}^A\tetrad^B{}_i,
\label{eq.54}
\end{split}
\ee
where in the second line the spin connection has been written explicitly in terms of the Lorentz matrices. 
Since the spin connection is independent from the tetrads the second line will not be needed in considering the variation of the spatial tetrads, but it will affect the variations in $\Lambda$.\\

In the following, to not repeat calculations unnecessarily we introduce the notion of the torsion tensor and Hamiltonian in the Weitzenböck gauge, respectively denoted by $\overset{w}{T}{}^A{}_{ij}$ and $\overset{w}{H}_\mathrm{TEGR}$. Note that we will omit Dirac deltas from our equations, unless they present spatial derivatives acting on them. From Eq. \eqref{eq.54} we get that
\be 
\begin{split}
\frac{\delta T^A{}_{ij}}{\delta \tetrad^C{}_k}&=\frac{\delta \overset{w}{T}{}^A{}_{ij}}{\delta \tetrad^C{}_k}+(\spinconnection^A{}_{Bi}\delta^B{}_C\delta_j^k-\spinconnection^A{}_{Bi}\delta^B{}_C \delta^k_i) \\
&=\frac{\delta \overset{w}{T}{}^A{}_{ij}}{\delta \tetrad^C{}_k}+2\spinconnection^A{}_{C[i}\delta_{j]}^k.
\label{eq.55}
\end{split}
\ee
Using this basic expression, we can write a relation between the variation with respect to the spatial tetrad of the Weitzenb\"{o}ck and the covariant Hamiltonian as
\be
\begin{split}
    \frac{\delta H_\mathrm{TEGR}}{\delta \tetrad^A{}_i}&=\frac{\delta \overset{w}{H}_\mathrm{TEGR}}{\delta \tetrad^A{}_i}+\shift^i\momenta_B{}^j\spinconnection^B{}_{Aj}-\lambda_{[AB]}\momenta_C{}^i\eta^{BC}+\lapse \momenta_C{}^{k}\spinconnection^C{}_{Bk}\frac{\delta \normalvector^B }{\delta \tetrad^A{}_i}\\
    &-\frac{\lapse\sqrt{\inducedmetric}}{\kappa}\frac{\partial \ {}^3 \mathbb{T} }{\partial T^C{}_{kl} }\spinconnection^C{}_{A[k}\delta^i_{l]}+2\shift^j\momenta_C{}^k\spinconnection^C{}_{A[k}\delta^i_{j]}\\
    &-2{}^{\mathcal{V}}\lambda^k\tetrad_B{}^l \spinconnection^B{}_{A[k}\delta^i_{l]} + {}^{\mathcal{A}}\lambda^{lk}\spinconnection^B{}_{A[k}\delta^i_{l]}\normalvector_B,
    \label{eq.56}
\end{split}
\ee
with the precaution that the result of the variation of $ {}^3 \mathbb{T} $ has to be considered for the torsion tensor in the covariant version, that is, including the spin connection.

We can obtain the same kind of relation for the variation in terms of the conjugate momenta of the spatial tetrad, which looks simpler and is given by
\be
\begin{split}
    \frac{\delta H_\mathrm{TEGR}}{\delta \momenta_A{}^i}=\frac{\delta \overset{w}{H}_\mathrm{TEGR}}{\delta \momenta_A{}^i}+\lapse\omega^A{}_{Bi}\normalvector^B+\shift^j\spinconnection^A{}_{Bi}\tetrad^B{}_j-\lambda_{[CB]}\eta^{BA}\tetrad^C{}_i.
    \label{eq.57}
\end{split}
\ee

An additional variation appears for the covariant Hamiltonian in terms of the conjugate momenta of the Lorentz matrices, which is
\be
\frac{\delta H_\mathrm{TEGR}}{\delta P^A{}_B}=-\lambda_{[AE]}\eta^{EC}\Lorentz_C{}^B.\label{eq.58}
\ee

Finally, we calculate the variation with respect to the the Lorentz matrices. For this the following identity will be useful
\be 
\frac{\delta \omega^C{}_{Di}}{\delta \Lorentz_A{}^B}=-\spinconnection^C{}_{Bi}\left(\Lorentz^{-1}\right)^A{}_D - \left(\Lorentz^{-1}\right)^A{}_D\delta^C_B \partial_i\delta
\label{eq.59}
\ee
Now we have all the necessary mathematical tools to compute Hamilton's equations for the covariant version of TEGR. \\

The Hamiltonian constraint is given by:
\be
\begin{split}
-{}^\alpha \dot{\pi}=\frac{\delta H}{\delta \alpha}=\frac{\kappa}{2\sqrt{\inducedmetric}} {}^{\mathcal{S}}{\momenta}_{ij} {}^{\mathcal{S}} \momenta^{ij} -\frac{3\kappa}{4\sqrt{\inducedmetric}}{}^{\mathcal{T}} \momenta {}^{\mathcal{T}} \momenta -\frac{\sqrt{\inducedmetric}}{2\kappa}{}^3 \mathbb{T}-\normalvector^A\partial_i \momenta_A{}^i  +\momenta_A{}^i\spinconnection^A{}_{Bi}\normalvector^B
\label{eq.60}
\end{split}
\ee
while the momentum constraint is given by
\be
-{}^\beta \dot{\pi}_i=\frac{\delta H}{\delta \beta^i}=-\tetrad^A{}_i \partial_j \momenta_A{}^j+\momenta_A{}^j\spinconnection^A{}_{Cj} \tetrad^C{}_i-\momenta_A{}^j T^A{}_{ji} \label{eq.61}
\ee
The variation with respect to the tetrad is
\begin{align}
    \begin{split}
        -\dot{\pi}_A{}^i= \dfrac{\delta H}{\delta \tetrad^{A}{}_{i} } &= \dfrac{\lapse \kappa}{2\sqrt{\inducedmetric}}\left(  \momenta_A{}^j \momenta_B{}^i \tetrad^B{}_j-\momenta_A{}^i \momenta_B{}^j \tetrad^B{}_j+ \inducedmetric^{il}\inducedmetric_{jk}\momenta_A{}^j\momenta_B{}^k \tetrad^B{}_l\right)
        +\shift^i\left(\momenta_B{}^j \spinconnection^B{}_{Aj} -\partial_j \momenta_A{}^j\right)\\
        &+\lambda_{[CA]}\momenta_B{}^i \eta^{BC} +\frac{\kappa}{\sqrt{\inducedmetric}}{}^{\mathcal{A}}\lambda_{[jk]}\momenta_A{}^j\inducedmetric^{ik} + {}^{\mathcal{V}}\lambda^l T^B{}_{kl}(\tetrad_{A}{}^k\tetrad_{B}{}^{i}+\normalvector_{A}\normalvector_{B}\gamma^{ki} )\\
        &+\lapse \normalvector_A \inducedmetric^{ik}\tetrad^B{}_k\left(\partial_j \momenta_B{}^j-\momenta_D{}^j \spinconnection^D{}_{Bj}\right)-{}^{\mathcal{V}}\lambda_j\frac{\kappa}{\sqrt{\inducedmetric}}\normalvector_A \inducedmetric^{ik}\tetrad^B{}_k \momenta_B{}^j -\dfrac12{}^{\mathcal{A}}\lambda^{lk}T^B{}_{kl}\normalvector_A \tetrad_B{}^i\\
        &+\frac{\lapse \tetrad_A{}^i}{2  }\left(-\frac{\sqrt{\inducedmetric}}{\kappa}{}^3 \mathbb{T}+\frac{\kappa}{2\sqrt{\inducedmetric}}\momenta_B{}^j\momenta_D{}^k\left(\tetrad^B{}_j\tetrad^D{}_k-\tetrad^B{}_k \tetrad^D{}_j-\inducedmetric_{jk}\inducedmetric^{ln}\tetrad^B{}_l\tetrad^D{}_n \right) \right)\\
        &-\frac{\kappa}{\sqrt{\inducedmetric}} {}^{\mathcal{V}}\lambda_j\tetrad_A{}^i\normalvector^B\momenta_B{}^j +{}^{\mathcal{V}}\lambda_j T^{B}{}_{kl}\tetrad_B{}^{k}(\tetrad_A{}^{j}\inducedmetric^{il} + \tetrad_{A}{}^{l} \inducedmetric^{ij} ) +\frac{\kappa}{\sqrt{\inducedmetric}} {}^{\mathcal{A}}\lambda_{[lj]} \inducedmetric^{kl}\tetrad_A{}^i\tetrad^B{}_k\momenta_B{}^j  \\
        &-\dfrac12{}^{\mathcal{A}}\lambda_{nj}T^B{}_{kl} \normalvector_B \left[\inducedmetric^{jl}(\tetrad_{A}{}^{n}\inducedmetric^{ik} + \tetrad_{A}{}^{k}\inducedmetric^{ni} ) + \inducedmetric^{nk}(\tetrad_{A}{}^{j} \inducedmetric^{il} + \tetrad_{A}{}^{l}\inducedmetric^{ji} ) \right]  \\
        &+\frac{\kappa}{\sqrt{\inducedmetric}}{}^{\mathcal{A}}\lambda_{[lj]} \pi_B{}^j \tetrad^B{}_k (\tetrad_{A}{}^{l}\inducedmetric^{ik} + \tetrad_{A}{}^{k}\inducedmetric^{li} ) \\
        &+ \frac{\kappa \lapse}{2\sqrt{\inducedmetric}}\eta_{AC}\left(\inducedmetric^{kl}\tetrad^B{}_l\tetrad^D{}_k\tetrad^C{}_j \momenta_B{}^i\momenta_D{}^j-\inducedmetric_{jk}\inducedmetric^{lm}\inducedmetric^{in}\tetrad^B{}_n\tetrad^D{}_l\tetrad^C{}_m\momenta_B{}^j\momenta_D{}^k \right) \\
        & +2\left(\shift^{[l}\momenta_B{}^{i]}-{}^{\mathcal{V}}\lambda^{[i} \tetrad_B{}^{l]}+\frac{1}{2} {}^{\mathcal{A}}\lambda^{[il]}\normalvector_B-\frac{\lapse\sqrt{\inducedmetric}}{\kappa}H_{CB}{}^{[mn][il]}T^C{}_{mn}\right)\spinconnection^B{}_{Al}\\
        &-2\partial_{l}\left(\shift^{[l}\momenta_A{}^{i]}-{}^{\mathcal{V}}\lambda^{[i} \tetrad_A{}^{l]} +\frac{1}{2} 
 {}^{\mathcal{A}}\lambda^{[il]}\normalvector_A -\frac{\lapse\sqrt{\inducedmetric}}{\kappa}H_{CA}{}^{[mn][il]}T^C{}_{mn}\right)\\
        &-\frac{\lapse \sqrt{\inducedmetric}}{\kappa}T^B{}_{kl}T^C{}_{mn}\left( \tetrad_{A}{}^{m}H_{CB}{}^{inkl} + \tetrad_{A}{}^{n}H_{CB}{}^{mikl} + \tetrad_{A}{}^{k}H_{CB}{}^{mnil} + \tetrad_{A}{}^{l}H_{CB}{}^{mnki} \right.\\
        &\left. + \normalvector_C \normalvector_A \inducedmetric^{i[m}\inducedmetric^{n][k}\tetrad_{B}{}^{l]}  + \normalvector_B \normalvector_A \tetrad_{C}{}^{[m}\inducedmetric^{n][k}\inducedmetric^{l]i} \right).
        \label{eq.62}
    \end{split}
 \end{align}
A new variation appears in the covariant formalism due to the inclusion of the Lorentz matrices as an additional canonical variable, therefore the time evolution of their associated canonical momenta are
\begin{align}
    \begin{split}
        -\dot{P}^B{}_A = \dfrac{\delta H}{\delta \Lambda_{B}{}^{A} } & = \frac{2\lapse \sqrt{\inducedmetric}}{\kappa}H_{CD}{}^{[ij][kl]}T^D{}_{kl}\left(\Lorentz^{-1} \right)^B{}_E \spinconnection^C{}_{Ai}\tetrad^E{}_{j} 
        \\&-\partial_{i}\left(\frac{2\lapse \sqrt{\inducedmetric}}{\kappa}H_{AD}{}^{[ij][kl]}T^D{}_{kl}\left(\Lorentz^{-1} \right)^B{}_E\theta^{E}_{j}\right)\\
        &-\left(\lapse \momenta_C{}^i\normalvector^E \left(\Lorentz^{-1} \right)^B{}_E \spinconnection^C{}_{Ai}-\partial_{i}\left(\lapse \momenta_C{}^i\normalvector^E \left(\Lorentz^{-1} \right)^B{}_E\right)\right)- \lambda_{CD} P^{[C}{}_{A} \eta^{D]B}\\
        &+2\left(\Lorentz^{-1} \right)^B{}_E\left(-{}^{\mathcal{V}}\lambda_k  \tetrad_C{}^{i}\inducedmetric^{jk}+\dfrac12 {}^{\mathcal{A}}\lambda^{ij}\normalvector_C\right)\spinconnection^C{}_{A[i}\tetrad^E{}_{j]}\\
        &-\partial_{i}\left(2\left(\Lorentz^{-1} \right)^B{}_E\left(-{}^{\mathcal{V}}\lambda_k  \tetrad_C{}^{[i}\inducedmetric^{j]k}+\dfrac12 {}^{\mathcal{A}}\lambda^{[ij]}\normalvector_C\right)\theta^{E}{}_{j}\right)\\
        &-\beta^i \left(\Lorentz^{-1} \right)^B{}_E \momenta_C{}^j \tetrad^E{}_j\spinconnection^C{}_{Ai}-\partial_{i}\left(\beta^i \left(\Lorentz^{-1} \right)^B{}_E \momenta_A{}^j \tetrad^E{}_j  \right).
        \label{eq.63}
    \end{split}
\end{align}
Variation in terms of all canonical momenta give
\begin{align}
    \dot{\alpha}=\frac{\delta H}{\delta {}^\alpha \pi}=- {}^{\alpha}\lambda, \label{eq.67}
\end{align}
\begin{align}
    \dot{\beta}^i=\frac{\delta H}{\delta {}^\beta\pi_i}=- {}^{\beta}\lambda^{i}, \label{eq.68}
\end{align}
\begin{align}
    \begin{split}
         \dot{\theta}^A{}_i=\frac{\delta H}{\delta \pi_A{}^i}&= \lapse\left( \frac{\kappa}{2\sqrt{\inducedmetric}}\left[ 2\momenta_B{}^j\tetrad^A{}_{[j}\tetrad^B{}_{i]} +\momenta_B{}^j \tetrad^A{}_k \tetrad^B{}_l\inducedmetric_{ij}\inducedmetric^{kl}\right]+\partial_{i}\normalvector^{A} +\normalvector^B \spinconnection^A{}_{Bi} \right)\\
         &+\shift^j\left(\tetrad^B{}_j \spinconnection^A{}_{Bi}-T^A{}_{ij} \right)+\partial_i\left(\beta^{j}\tetrad^{A}_{j}\right)+{}^{\mathcal{V}}\lambda_i \frac{\kappa \normalvector^A}{\sqrt{\inducedmetric}} +{}^{\mathcal{A}}\lambda_{[ij]}\frac{\kappa \inducedmetric^{kj}\tetrad^A{}_k}{\sqrt{\inducedmetric}}  +\lambda_{[BC]}\eta^{AB}\tetrad^C{}_i,
         \label{eq.69}
    \end{split}
\end{align}
\begin{align}
    \begin{split}
        \dot{\Lambda}_A{}^B&=\frac{\delta H}{\delta P^A{}_B}=-\lambda_{[AD]}\Lambda_C{}^B\eta^{CD}. \label{eq.70}
    \end{split}
\end{align}
Just like in the previous case in the pure-tetrad TEGR, these equations have to be complemented with the primary constraints  \eqref{Calph},\eqref{Cbeth},\eqref{eq:CovConstraints},\eqref{Cvek} and \eqref{Casy} that set conditions on the initial data. The new sets of equations \eqref{eq.63} and \eqref{eq.70}, are obtained from the introduction of the Lorentz matrices in the formalism \cite{Golovnev:2021omn}, and in particular \eqref{eq.70} shows that the time evolution of the Lorentz matrices is arbitrary and determined by Lagrange multipliers.

\section{Discussion and comparison with previous works}
\label{sec:disc}

The comparison of our results with already existing work in the literature complicates due to the variety of different ways for tackling the Hamiltonian formalism for TEGR. For instance, in a recent review \cite{Blixt:2020ekl}, there are summarized at least five different formalisms, methods and notation that have been used in the literature. \\

One of the first works in the literature discussing the time evolution of the tetrad and momenta is in \cite{MTh,Cheng:1988zg} and partially in \cite{Blagojevic00}. In Appendix \ref{app:C} we show that the results presented in \cite{Blagojevic00} are consistent with ours. The results of \cite{Cheng:1988zg} apply only for the class of theories known as one-parameter teleparallel gravity, not TEGR. Instead of computing explicitly the time evolution of variables through its Poisson bracket with the Hamiltonian, they use Hamilton's equations to analyze the time evolution of constraints and canonical variables. They arrive to the important conclusion that one-parameter teleparallel gravity (OPTG) has six physical degrees of freedom, a result that has not yet been confirmed independently by other authors. Thus, it would be an interesting future direction to extend this work to OPTG and confirm or disprove their results, especially because the calculations are very lengthy. The most elaborated calculations are in \cite{MTh}, although we have found some typos and even seemingly some mistakes in the extensive calculations. However, the most simple perturbations at the lowest order around a Minkowski background only propagate three degrees of freedom, while it has been shown that additional degrees of freedom propagate at higher orders \cite{BeltranJimenez:2019nns}. This suggests that the conclusion of six physical degrees of freedom is plausible and that they could manifest perturbatively. Another exhaustive reference where the Hamiltonian structure of TEGR and generic NGR is computed is \cite{Mitric:2019rop}. Although the time evolution for the tetrad and their momenta are not specified, it can be used as a starting point for their derivation.\\

In relation with the work developed in \cite{Capozziello:2021pcg}, there are several points where our approaches differ. The authors chose a $3+1$ split of Latin indices in temporal and spatial parts as  $A=(\tilde{0},\tilde{i})$ in the same way that spacetime indices  $\mu=(0,i)$ are commonly split in the literature. They conclude that $e^{\tilde{0}}{}_{i}=0$ (after their Eq.(57)), but with some flaws in the reasoning for arriving to this expression. They claim correctly that the induced metric can be written in terms of the components of this tetrad splitting as $e^{\tilde{0} }{}_{i} e^{\tilde{0} }{}_{j} \eta_{\tilde{0}\tilde{0} } + e^{\tilde{i} }{}_{i} e^{\tilde{j} }{}_{j} \eta_{\tilde{i}\tilde{j} } = \gamma_{ij}$. However, immediately after that they assert instead that $ e^{\tilde{i} }{}_{i} e^{\tilde{j} }{}_{j} \eta_{\tilde{i}\tilde{j} } = \gamma_{ij}$ (which was explained in \cite{Blixt:2020ekl} to be an unnecessary choice), and conclude that $e^{\tilde{0} }{}_{i} $ must be zero, instead of admitting that this is a fixing of the tetrad components. As explained in \cite{Blixt:2020ekl} the form of the tetrad they adopted fixes a Lorentz gauge. If their work should be extended to more general teleparallel theories it may enforce a nontrivial spin connection. Even in the case of TEGR it might turn out that this gauge choice is unfavorable for numerical relativity, or that a boundary term needs to be added to get a well-defined ADM mass for instance (in the context of the role of boundary terms and the inertial frame see \cite{Gomes:2022vrc}). \\

In addition to the advantage of our formalism being gauge independent and covariant, there seems to be a fundamental advantage of using the covariant phase-space variables from the Hamiltonian rather than fields like extrinsic curvature in tetrad theories of gravity for the following reason. When writing the evolution equations of the tetrad field the antisymmetric fields generically evolve. What happens in metric theories of gravity is that the field equations are used to rewrite the symmetric evolution equations in a desired form. The problem that occurs in TEGR is that there are no antisymmetric field equations while there is antisymmetric evolution \cite{Capozziello:2021pcg}. Thus, it is unclear what one should do with the antisymmetric evolution equations. In contrast to the field equations in Lagrangian variables Hamilton's equations have antisymmetric equations (see \eqref{Cvek} and \eqref{Casy}). So this kind of difficulty is avoided in our approach. 

\section{Conclusions}
\label{sec:conc}

In this work we present Hamilton's equations for the teleparallel equivalent of general relativity, following the Hamiltonian approach introduced in \cite{Blixt19} using the VAST decomposition of the momenta and tetrad velocities. We also present a novel derivation of Hamilton equations for the covariant version of TEGR, with an arbitrary spin connection depending on Lorentz matrices. To our knowledge the case of Hamilton's equations for TEGR in the Weitzenböck gauge has only been performed in \cite{MTh}, although the formalism is slightly different from ours. The results for the Weitzenböck gauge are given by equations \eqref{eq.42}-\eqref{eq.49} in Sec.\ref{sec:HeqsW}. Equations \eqref{eq.60}-\eqref{eq.70} in Sec. \ref{sec:Heqscov} are Hamilton's equations for the covariant formulation.\\

As expected, there are two main differences between (covariant) TEGR Hamilton's equations derived here and Hamilton's equations of GR in its standard formulation (see Appendix \ref{app:B}). The first difference is related to which fields are treated as canonical. Tetrads have six additional independent components compared to the metric. The spin connection (or rather the Lorentz matrices) further introduces six more additional independent components \cite{Golovnev:2021omn}. There exists Hamilton's equations for all of the canonical fields and their conjugate momenta. Thus, one main difference is that Hamilton's equations for TEGR are more numerous than those in the standard formulation of GR. Note, however, that the EH action can straightforwardly be rewritten in terms of tetrads and we would get a structure more reminiscent to the case of TEGR in the Weitzenböck gauge. The second difference is related to the boundary term. It seems that TEGR Hamilton's equations are more lengthy than the standard ones. However, we would like to point out that the standard formulation \cite{Arnowitt:1962hi} is different from the EH action by the York-Gibbons-Hawking boundary term \cite{York72,Gibbons77}. An interesting future direction could be to investigate if the gauge conditions denoted as ``inertial frame'' defined in \cite{Gomes:2022vrc} would give any kind of an advantage over the standard formulation.\\

Despite those differences, the counting of degrees of freedom is the same for (covariant) TEGR and GR, as expected. From the $16 + 6$ independent components of the tetrad $\theta^{A}{}_{\mu}$ and the Lorentz matrices $\Lambda^{A}{}_{B}$ \cite{Golovnev:2021omn}, there are $4+6+6$ primary constraints and $4$ secondary constraints (Hamiltonian and momenta) which are all first class, therefore we are left with $24-20=4$ degrees of freedom in the phase space, consequently two physical degrees of freedom. Proving that all constraints are first class is beyond the scope of our paper, but it has been proved elsewhere, leading to the same counting of degrees of freedom as presented above. We also do not compute the preservation over time of primary constraints, nor do we compute gauge generators or consider nonlinear extension of TEGR, where the issue of the counting of degrees of freedom overly complicates. However, we hope that our work can be a useful contribution in this direction.\\

The tetrad formulation of Einstein equations for numerical relativity has been considered in, for instance, Ref. \cite{Buchman:2003sq}. Here it is asserted that since tetrad frames are natural for measuring observable physical quantities, as for instance they are tied to the flow of a fluid, in either cosmology or interior metrics of rotating stars, then they can be used even for vacuum black hole spacetimes. Therefore, it is interesting to check if the same conclusions can be drawn when working in the tetrad formalism in TEGR. In particular, it would be pertinent for future work to study the hyperbolicity of the 3+1 equations of motion in TEGR for the tetrad field, and if it can be of help for the improvement of efficiency of numerical relativity.\\

In summary, Hamilton's equations look much more lengthy in (covariant) TEGR formulation compared to the standard metric formulation. Though, in addition to setting up the path for deriving Hamilton's equations for modified teleparallel theories, it will still need to be investigated if this formulation have advantages for numerical relativity. The conjugate momenta to the metric was identified to be a part of the conjugate momenta of the tetrad using the VAST decomposition. However, the converse cannot be obtained due to the fact that the metric has more symmetries than the tetrad. It was also noted that Hamilton's equations were affected by the boundary term. Lastly, we note that the formulation in this article does not have undetermined antisymmetric evolution equations (here they are given through Lagrange multipliers that trivialize in Lagrangian field variables) thus this approach seems to have an advantage over the approach performed in \cite{Capozziello:2021pcg}.

\section*{Acknowledgments}

The authors thank Jose Beltr\'an-Jim\'enez, Alejandro Jim\'enez-Cano, Tomi Koivisto,  Laur J\"{a}rv, and an anonymous referee for their helpful comments, corrections and criticism on a previous version of this manuscript. The authors are grateful to Bivudutta Mishra for organizing “27th International Conference of
International Academy of Physical Sciences on Advances in Relativity and Cosmology” which was the starting point for the collaboration established by the authors. L. P. is a beneficiary of the Dora Plus Program, organized by the University of Tartu also supported by Department of Science and Technology (DST), Government of India, New Delhi for awarding INSPIRE fellowship (File No. DST/INSPIRE Fellowship/2019/IF190600) to carry out the research work as Senior Research Fellow. It is L. P's pleasure to thank her supervisor, professor. Bivudutta Mishra, for his continuous support. L. P and M. J. G. have  been supported by the European Regional Development Fund CoE program TK133 “The Dark Side of the Universe”. M. J. G has been supported by the Estonian Research Council Grant No. MOBJD622.  D. B would like to acknowledge the contribution of the COST Action CA18108.

\appendix

\section{Useful identities}
\label{app:A}

Some simple variation formula of canonical fields are
\be 
\frac{\delta \tetrad }{\delta \tetrad^{A}{}_{\mu} } = \tetrad \cotetrad^{\mu}_{A},
\ee 

\be 
\frac{\delta \cotetrad }{\delta \tetrad^{A}{}_{\mu} } = -\cotetrad \cotetrad_A{}^{\mu},
\ee 

\be
\frac{\delta \left(\Lambda^{-1}\right)^E{}_D}{\delta \Lambda_A{}^B}=-\left(\Lambda^{-1} \right)^E{}_B\left(\Lambda^{-1} \right)^A{}_D.
\ee
Up to now, all previous expression hold in the phase space quantities before performing the 3+1 decomposition in the tetrad. The following relations hold for the components of the 3+1 decomposed tetrad $\tetrad^{A}{}_{i}$, the induced metric $\inducedmetric_{ij}$, its inverse $\inducedmetric^{ij}$, and the vector $\xi_A$:
\begin{align}
\frac{\delta \theta_{B}{}^{j}}{\delta \theta^{A}{}_{i}}&=-\theta_{A}{}^{j}\theta_{B}{}^{i} -\normalvector_A \normalvector_B \inducedmetric^{ij},\\
\frac{\delta\gamma^{ij}}{\delta\theta^{C}{}_{m}}&=-\theta_{C}{}^{i}\gamma^{mj}-\theta_{C}{}^{j}\gamma^{im},\\
\frac{\delta \normalvector^{A}}{\delta \theta^{C}{}_{m}}&=-\normalvector_{C}\theta^{A}{}_{j}\gamma^{jm},\\
\frac{\delta\gamma_{ij}}{\delta\theta^{D}{}_{n}}&=\eta_{AD}\left(\delta^{n}_{i}\theta^{A}{}_{j}+\delta^{n}_{j}\theta^{A}{}_{i}\right),\\
    \frac{\delta\sqrt{\gamma}}{\delta\theta^{C}{}_{m}}&=\sqrt{\gamma}\gamma^{im}\theta^{A}{}_{i}\eta_{AC},\\
    \frac{\delta}{\delta\theta^{C}{}_{m}}\left(\frac{1}{\sqrt{\gamma}}\right)&=\frac{-1}{\sqrt{\gamma}}\gamma^{im}\theta^{A}{}_{i}\eta_{AC},\\
    \dfrac{\delta T^{A}{}_{ij} }{\delta T^{C}{}_{mn} } &= \dfrac12 \delta^{A}_{C}(\delta^m_i\delta^n_j - \delta^m_j \delta^n_i ),
 \end{align}   
\begin{align}    
\dfrac{\delta {}^3{\mathbb T} }{\delta T^{C}{}_{mn} }& = H_{CB}{}^{mnkl} T^{B}{}_{kl} + T^{A}{}_{ij} H_{AC}{}^{ijmn} = 2H_{AC}{}^{ijmn} T^{A}{}_{ij},
\end{align}
\begin{equation}
\dfrac{\delta T^{C}{}_{ij} }{\delta \omega^{D}{}_{Ek} } = \delta^{C}_{D}( \delta^{k}_{i} \tetrad^{E}_{j} - \delta^{k}_{j} \tetrad^{E}_{i} ),
\end{equation}
\begin{equation}
\dfrac{\delta T^{C}{}_{ij} }{\delta \Lambda_{B}{}^{A} }  =  (\Lambda^{-1})^{B}{}_{E}\left( \tetrad^{E}_j \omega^{C}{}_{Ai} - \tetrad^{E}_i \omega^{C}{}_{Aj}- \delta^{C}_{A}\tetrad^{E}_j\partial_i\delta + \delta^{C}_{A}\tetrad^{E}_{i}\partial_j\delta \right),
\end{equation}
\begin{equation}
\frac{\delta T^{A}{}_{ij}}{\delta \theta^{C}{}_{m}} = \delta^{A}_{C}\left(\delta^{m}_{j} \partial_{i}\delta  - \delta^{m}_{i} \partial_{j}\delta \right)+2\omega^{A}{}_{C[i}\delta^{m}{}_{ j]},
\end{equation}
\begin{align}
    \begin{split}
         \dfrac{\delta H_{AB}{}^{ijkl} }{\delta \tetrad^{C}{}_{m} } &= \tetrad^{i}{}_{C} H_{AB}{}^{mjlk} + \tetrad^{j}{}_{C} H_{AB}{}^{mikl} + \tetrad^{k}{}_{C}H_{AB}{}^{ijlm}+\tetrad^{l}{}_{C}H_{AB}{}^{ijmk}-\frac{1}{2}\normalvector_B\normalvector_C\inducedmetric^{m[i}\inducedmetric^{j][k}\tetrad_A{}^{l]}\\
         &-\frac{1}{2}\normalvector_A\normalvector_C \tetrad_B{}^{[i}\inducedmetric^{j][k}\inducedmetric^{l]m}+\normalvector_A\normalvector_C\inducedmetric^{m[i}\inducedmetric^{j][k}\tetrad_B{}^{l]}+\normalvector_B\normalvector_C\tetrad_A{}^{[i}\inducedmetric^{j][k}\inducedmetric^{l]m}.
    \end{split}
\end{align}

\section{Hamilton's equations in General Relativity}
\label{app:B}
The derivation of the Hamiltonian in TEGR has more intricacies than the standard EH formulation,
since the tetrad, our dynamical variable, has more free functions that have to be dealt with compared  to the metric. In order to compare TEGR and EH Hamiltonians, it will be instructive to take a look at the main results for the EH action,
\begin{align}
    S_{\mathrm{EH}}=\frac{1}{2\kappa}\int \mathrm{d}^4 x R,
\end{align}
which can be found in classical references \cite{Arnowitt:1962hi,Wald:1984rg}. Those are in done in metric formulation, in tetrad formulation the Hamiltonian is presented in \cite{Hinterbichler:2012cn}. However, we are not aware of any reference presenting Hamilton's equations for tetrad EH formulation.
The 3+1 split of the metric goes as
\begin{equation}
    g_{\mu\nu} = \left( \begin{array}{cc}
        -\alpha^2 + \beta^{i}\beta^{j}\gamma_{ij} & \beta_{i}  \\
        \beta_{j} & \gamma_{ij}
    \end{array} \right), \hspace{2cm} g^{\mu\nu} = \dfrac{1}{\alpha^2 }\left( \begin{array}{cc}
        -1 & \beta^{i} \\
        \beta^{j} & \alpha^2 \gamma^{ij} - \beta^{i}\beta^{j}
    \end{array}\right).
\end{equation}

The Lagrangian density of EH with a boundary term already discarded\footnote{Note that in TEGR action such a boundary term is not necessary} is written as
\begin{equation}
    \mathcal{L} = \sqrt{\gamma} \alpha[ {}^{(3)}R + K_{ij} K^{ij} - K^2 ],
\end{equation}
where the extrinsic curvature $K_{ij}$ is related to the time derivative of the induced metric $\gamma_{ij}$ as
\begin{equation}
    K_{ij} = \dfrac{1}{2\alpha} [ \dot{\gamma}_{ij} - D_i \beta_j - D_j \beta_i ].
\end{equation}
The momenta canonically conjugate to the induced metric $\gamma_{ij}$ are
\begin{equation}
\label{momGR}
    \pi^{ij} = \dfrac{\partial \mathcal{L} }{\partial \dot{\gamma}_{ij} } = \sqrt{\gamma}( K^{ij} - K \gamma^{ij})
\end{equation}
By rewriting the Lagrangian density in terms of these momenta, the canonical Hamiltonian density is obtained:
\begin{align}
\begin{split}
    \mathcal{H} = & \pi^{ij}\dot{\gamma}_{ij} - \mathcal{L} \\
    = & -\sqrt{\gamma} \alpha {}^{(3)}R + \alpha \sqrt{\gamma} \left( \pi^{ij}\pi_{ij} - \dfrac12 \pi^2 \right) + 2\pi^{ij}D_i \beta_j \\
    = & \sqrt{\gamma}\left( \alpha\left[ -{}^{(3)}R + \pi^{ij}\pi_{ij}/\gamma - \dfrac12 \pi^2/\gamma \right] -2\beta_j[D_i(\pi^{ij}/\sqrt{\gamma} ) ] + 2D_i(\beta_j \pi^{ij}/\sqrt{\gamma} ) \right)
    \end{split}
\end{align}
where $\pi^{k}_{k}=\pi$. The last term is a boundary term and it is discarded (analogous to what was done in this work for TEGR discarding $\partial_i \left[\momenta_A{}^i\tetrad^A{}_0 \right]$),  which is essential, since the boundary term contains spatial derivatives of the  shift and would complicate its interpretation as  Lagrange multipliers, and otherwise it would manifest in Hamilton's equations with nontrivial dynamics. The same would happen with the lapse function in TEGR if we did not discard the boundary term. Together with the momenta \eqref{momGR} we obtain primary constraints coming from the absence of $\dot{\alpha}$ and $\dot{\beta}^{i}$ in the Lagrangian, which we denote as ${}^{\alpha} \pi_{EH} \approx 0$ and ${}^{\beta} \pi^{j}_{EH} \approx 0$, which should not be confused with ${}^\alpha \pi$ and ${}^\beta \pi^j$ which were defined with the TEGR Lagrangian. If considering lapse and shift as canonical variables from the beginning (before discovering their trivial role as Lagrange multipliers), then we must include these primary constraints with the corresponding Lagrange multipliers $({}^{\alpha} \lambda, {}^{\beta} \lambda_i )$ in the primary Hamiltonian, which becomes 
\begin{equation}
\mathcal{H}_p = \mathcal{H} - {}^{\alpha} \lambda {}^{\alpha} \pi_{EH} - {}^{\beta} \lambda_i {}^{\beta} \pi^{i}_{EH}.
\end{equation}
The variation of the primary Hamiltonian $H = \int d^3x \mathcal{H}_p$ with respect to $\alpha$ and $\beta_i$ yields, respectively, the equations
\begin{equation}
    {}^\lapse\dot{\pi}_{\mathrm{EH}}=\frac{\delta H}{\delta \lapse}=\sqrt{\inducedmetric}\left(-{}^{(3)}R + \dfrac{1}{\gamma}\left( \pi^{ij} \pi_{ij} - \frac{1}{2}\pi^2 \right)\right) = 0.
    \label{Heq_lapse}
\end{equation}
\begin{equation}
   {}^{\beta}\dot{\pi}_{\mathrm{EH}}^j=\frac{\delta H}{\delta \shift_j}= -2\sqrt{\inducedmetric}D_i( \pi^{ij}/\sqrt{\gamma}) = 0,
    \label{Heq_shift}
\end{equation}
These are identified as constraints for the initial values of momenta and induced metric, also known as Hamiltonian and momenta constraints, respectively. The absence of $\dot{\lapse}$ and $\dot{\shift}^i$ in the Lagrangian translates as the following Hamilton's equations
\begin{align}
    \dot{\lapse}=\frac{\delta H}{\delta {}^{\lapse}\momenta }= - {}^{\alpha} \lambda ,\\
     \dot{\shift}^i=\frac{\delta H}{\delta {}^{\shift}\momenta^i }= - {}^{\beta} \lambda_i,
\end{align}
which state that the time evolution of lapse and shift is determined by arbitrary Lagrange multipliers. Finally, the  dynamical equations obtained from the Hamiltonian, that is, Hamilton's equations for the induced metric and the momenta, are respectively,
\begin{equation}
\label{gammatimeev}
\dot{\gamma}_{ij} = \dfrac{\delta H}{\delta \pi^{ij} } = 2\gamma^{-1/2} \alpha\left( \pi_{ij} - \dfrac12 \gamma_{ij}\pi \right) + 2D_{(i}\beta_{j)},
\end{equation}
and
\begin{equation}
    \begin{split}
        \dot{\pi}^{ij} = & -\dfrac{\delta H}{\delta \gamma_{ij} } = -\alpha \sqrt{\gamma}\left( {}^{(3)}R^{ij} - \dfrac12 {}^{(3)}R \gamma^{ij} \right) + \dfrac12 \alpha \gamma^{-1/2} \gamma^{ij} \left( \pi^{kl}\pi_{kl} - \dfrac12 \pi^2 \right) \\
        &-2 \alpha \gamma^{-1/2} \left( \pi^{ik}\pi_{k}{}^{j} - \dfrac12 \pi \pi^{ij} \right)   + \sqrt{\gamma} (D^{i}D^{j}\alpha - \gamma^{ij} D^{k}D_{k}\alpha ) + \sqrt{\gamma} D_k( \gamma^{-1/2} \beta^k \pi^{ij} ) \\
        &- 2\pi^{k(i} D_{k}\beta^{j)}.
    \end{split}
\end{equation}

\section{Hamilton's equations in covariant TEGR (previous partial results)}
\label{app:C}
 In \cite{Blagojevic00} the Hamiltonian analysis for covariant TEGR was performed by imposing the teleparallel condition $R^{AB}{}_{\mu\nu}=0$ by means of Lagrange multipliers $\lambda_{AB}{}^{\mu\nu}$ and the addition of the term $\lambda_{AB}{}^{\mu\nu}R^{AB}{}_{\mu\nu}$ in the TEGR Lagrangian. In other words, the theory can be seen as a special case of Poincaré gauge theory, where curvature vanishes and torsion appears in the action in the well-known form of TEGR. Before presenting their results, some explanation regarding the difference in the formulation is in order (further details can be found in \cite{Blixt:2020ekl}). The canonical variables in \cite{Blagojevic00} are $(\theta^A{}_\mu, \omega^A{}_{B\mu},\lambda_{AB}{}^{\mu\nu})$, which differs from our choice $(\lapse,\shift^i, \theta^A{}_i,\Lorentz^A{}_B)$. This means that some of their equations will appear to be different, even though the set of equations are equivalent. Instead of introducing indices that run on the three-dimensional hypersurface of constant time slices $\Sigma$ they instead use projectors. A vector projected to the normal vector is denoted $V_\bot=\normalvector^A V_A$ and it is also defined $V_{\Bar{A}}=V_A-\normalvector_A V_\bot$. They define lapse $N$ and shift $N^i$ as $N=\normalvector_C \tetrad^C{}_0=\lapse$ and $N^i=\cotetrad_{\Bar{C}}{}^i \tetrad^C{}_0=\beta^i $ which shows that our definitions coincide.  The different set of canonical variables also implies a different set of conjugate momenta. However, $\pi_A{}^i$ coincides with our expression except for the formulation details using projectors. Instead of our conjugate momenta with respect to the Lorentz matrices they have additional sets of canonical momenta $\pi_{AB}{}^\mu:=\frac{\partial L}{\partial \dot{\omega}^{AB}{}_\mu}$ and $\pi^{AB}{}_{\mu\nu}:=\frac{\partial L}{\partial \dot{\lambda}_{AB}{}^{\mu\nu} }$. 
 
 \subsection{The Hamiltonian}
 
 They find the expression for the total Hamiltonian to be \footnote{This coincides with our primary Hamiltonian (not total, as the authors denote it in their paper). Technically, the total Hamiltonian corresponds to the primary Hamiltonian once the Lagrange multipliers have been solved and replaced back, see for instance \cite{Brown:2022gha}.}
 \begin{align}
     \begin{split}
         \mathcal{H}=\mathcal{H}_c+\frac{1}{2}\Bar{u}_{AB}{}^{0i}\pi^{AB}{}_{0i}+u^A{}_0\pi_A{}^0+\frac{1}{2}u^{AB}{}_0 \pi_{AB}{}^0+\frac{1}{4}u_{AB}{}^{ij}\pi^{AB}{}_{ij}+\frac{1}{2}u^{AC}\phi_{AC},
     \end{split}
     \label{eq:BlagHp}
 \end{align}
 with the canonical Hamiltonian defined as [second unnumbered equation after (3.5)]
 \begin{align}
     \begin{split}
         \mathcal{H}_c:=\pi_A{}^\mu\dot{\tetrad}^A{}_\mu+\momenta_{AB}{}^\mu \dot{\spinconnection}^{AB}{}_\mu-\sqrt{g} \mathcal{L}.
     \end{split}
 \end{align}
 Note that due to the different formulation, the canonical Hamiltonian is defined slightly different from our case. This is since instead of treating the Lorentz matrices as canonical variables, the spin connection is in this case treated as canonical. Explicitly, the canonical Hamiltonian is found here to be 
 \begin{align}
     \mathcal{H}_c=\lapse \mathcal{H}_\bot +\shift^i  \mathcal{H}_i-\frac{1}{2}\spinconnection^{AB}{}_0 \mathcal{H}_{AB}+\partial_i D^i, 
 \end{align}
 where the definitions of  $\mathcal{H}_\bot$, $\mathcal{H}_i$, $\mathcal{H}_{AB}$, $D^i$ defined in \cite{Blagojevic00} give the explicit Hamiltonian
 \begin{align}
     \begin{split}
         H_c&=\lapse\left(\hat{\pi}_\mu{}^{\Bar{A}}T^\mu{}_{\bot \Bar{A}}-\sqrt{\inducedmetric}\frac{1}{2\kappa}\left(\frac{1}{4}T_{ABC}T^{ABC}+\frac{1}{2}T_{ABC}T^{BAC}-T_A T^A \right)-\normalvector^A\nabla_i \momenta_A{}^i \right)\\
         &-\alpha\sqrt{\inducedmetric} \lambda_{AB}{}^{\Bar{C}\Bar{D}}R^{AB}{}_{\Bar{C}\Bar{D}} +\shift^i\left(\momenta_C{}^j  T^C{}_{ij}-\tetrad^C{}_i \nabla_j \momenta_C{}^j +\frac{1}{2}\momenta_{AB}{}^\beta R^{AB}{}_{ij} \right)\\
         &-\frac{1}{2}\spinconnection^{AB}{}_0 \left(2\momenta_{[A}{}^i \tetrad_{B]i}+\nabla_i \momenta_{AB}{}^i \right)+\partial_i \left(\tetrad^C{}_0 \momenta_C{}^i +\frac{1}{2}\spinconnection^{AB}{}_0 \momenta_{AB}{}^i \right).
     \end{split}
     \label{eq:BlagHc}
 \end{align}

\subsection{Poisson brackets}

The evolution of the canonical fields can equivalently be calculated with the help of Poisson brackets using the identity
\begin{align}
    \dot{A}=\{A,H\}.
\end{align}
In \cite{Blagojevic00} the following Poisson brackets were calculated using \eqref{eq:BlagHc}
\begin{align}
    \begin{split}
        \{\tetrad^A{}_i,\mathcal{H}'_{CD} \}=\delta^A_C \tetrad_{Di}\delta-(C \longleftrightarrow D),
    \end{split}
\end{align}
\begin{align}
    \begin{split}
        \{\tetrad^A{}_i, \mathcal{H}'_j \}=\nabla_i \tetrad^A{}_j \delta -\tetrad^A{}_j \partial _i \delta= T^A{}_{ji} \delta+\nabla_i \left(\tetrad^A{}_j \delta  \right),
    \end{split}
\end{align}
\begin{align}
    \begin{split}
        \{\tetrad^A{}_i , \mathcal{H}'_\bot \}=\kappa \left(\tetrad_{Ci} P_T^{\Bar{A}\Bar{C}}-\frac{1}{6}\tetrad^A{}_i P^{\Bar{C}}{}_{\Bar{C}} \right)\delta +\nabla_i \left(\normalvector^A \delta  \right),
    \end{split}
\end{align}
where $\delta$ is a shorthand notation for $\delta(\textbf{x}-\textbf{x}')$.
From this they conclude that 
\begin{align}
    \begin{split}
        \nabla_0 \tetrad^A{}_i=\nabla_i \tetrad^A{}_0 +\shift^i T^A{}_{ij}+\lapse \tetrad_{Cj}\frac{\kappa}{\sqrt{\inducedmetric}}\left(\hat{\pi}^{(\Bar{A}\Bar{C})}-\frac{1}{2}\eta^{\Bar{A}\Bar{C}}\hat{\pi}^{\Bar{B}}{}_{\Bar{B}} \right)+\tetrad_{Ci}\left(\normalvector^A u^{\bot \Bar{C}}+u^{\Bar{A}\Bar{C}}\right).
    \end{split}
\end{align}
When $\nabla \equiv \partial $ the terms containing neither momenta nor Lagrange multipliers coincide with our result. The terms that involve the conjugate momenta are also consistent with our findings. The parts including Lagrange multipliers are a bit more tricky to compare. Note that \cite{Blagojevic00} combines our 3+3 primary constraints into six primary constraints
\begin{align}
    \phi_{AB}=\pi_{A\Bar{B}}-\pi_{B\Bar{A}}+a \nabla_i\left( \epsilon^{0ijk}_{ABCD}\tetrad^C{}_j \tetrad^D{}_k\right),
\end{align}
which are essentially equivalent, up to changing the definition of $u^{AB}$ compared to the definitions in our manuscript $({}^\mathcal{V}\lambda_i,{}^\mathcal{A}\lambda_{ij})$. Note that here the factors $\sqrt{\inducedmetric}$ and $\kappa$ are not present (like in our work); nonetheless the expressions are consistent, since those are overall factors that in our formalism could have been absorbed by the Lagrange multipliers. Otherwise $u_{\Bar{A}\Bar{C}}$ coincides with ${}^{\mathcal{A}}\lambda_{[ij]}$ and $u^{\bot \Bar{C}}\tetrad_{Ci}$ coincides with ${}^{\mathcal{V}}\lambda_i$, which can be proved in a similar way as it was done in \cite{Blixt:2020ekl}. In summary, their results for the tetrad time evolution are consistent with ours. 

There are some other interesting calculations in \cite{Blagojevic00}, however, they do not correspond to the canonical variables considered in this article. The reader is, thus, referred to the original article for more details.

\end{document}